\documentclass[aps,prl,twocolumn,showpacs,floatfix,superscriptaddress,noshowpacs]{revtex4-1}
\usepackage{graphicx}
\usepackage{amsmath}
\usepackage{amssymb}
\usepackage{color}
\usepackage{float}
\usepackage[colorlinks=true,citecolor=blue,linkcolor=blue]{hyperref}

\usepackage[normalem]{ ulem }
\usepackage{soul}

\begin{document}

\flushbottom

\title{Low frequency imaginary impedance at the superconducting transition of 2H-NbSe$_2$} 

\author{David Perconte}
\affiliation{Laboratorio de Bajas Temperaturas y Altos Campos Magn\'eticos, Departamento de F\'isica de la Materia Condensada, Instituto Nicol\'as Cabrera and Condensed Matter Physics Center (IFIMAC), Universidad Aut\'onoma de Madrid, E-28049 Madrid,
Spain}

\author{Samuel Ma\~nas-Valero} 
\affiliation{Instituto de Ciencia Molecular (ICMol), Universidad de Valencia, Catedr\'atico Jos\'e Beltr\'an 2, 46980 Paterna, Spain}

\author{Eugenio Coronado} 
\affiliation{Instituto de Ciencia Molecular (ICMol), Universidad de Valencia, Catedr\'atico Jos\'e Beltr\'an 2, 46980 Paterna, Spain}

\author{Isabel Guillam\'on}
\affiliation{Laboratorio de Bajas Temperaturas y Altos Campos Magn\'eticos, Departamento de F\'isica de la Materia Condensada, Instituto Nicol\'as Cabrera and Condensed Matter Physics Center (IFIMAC), Universidad Aut\'onoma de Madrid, E-28049 Madrid,
Spain}
\affiliation{Unidad Asociada de Bajas Temperaturas y Altos Campos Magn\'eticos, UAM, CSIC, Cantoblanco, E-28049 Madrid, Spain}

\author{Hermann Suderow}
\affiliation{Laboratorio de Bajas Temperaturas y Altos Campos Magn\'eticos, Departamento de F\'isica de la Materia Condensada, Instituto Nicol\'as Cabrera and Condensed Matter Physics Center (IFIMAC), Universidad Aut\'onoma de Madrid, E-28049 Madrid,
Spain}
\affiliation{Unidad Asociada de Bajas Temperaturas y Altos Campos Magn\'eticos, UAM, CSIC, Cantoblanco, E-28049 Madrid, Spain}

\begin{abstract}
The superconducting transition leads to a sharp resistance drop in a temperature interval that can be a small fraction of the critical temperature T$_c$. A superconductor exactly at T$_c$ is thus very sensitive to all kinds of thermal perturbations, including the heat dissipated by the measurement current. We show that the interaction between electrical and thermal currents leads to a sizeable imaginary impedance at frequencies of order of tens of Hz at the resistive transition of single crystals of the layered material 2H-NbSe$_2$. We explain the result using models developed for transition edge sensors. By measuring under magnetic fields and at high currents, we find that the imaginary impedance is strongly influenced by the heat associated with vortex motion and out-of-equilibrium quasiparticles.
\end{abstract}

\maketitle

\section{Introduction}

Since the discovery of superconductivity by H. Kamerling-Onnes over a century ago, the resistive transition continues to fascinate researchers, in spite of being now a routine measurement in many laboratories all over the world. The transition is often very sharp, which allows to build extremely sensitive thermometers from superconductors stabilized at T$_c$. These are called transition edge sensors (TES) and are used in X-ray and $\gamma$-ray detection\cite{EnssBook,Ullom_2015,Redfern2002}.

However, this implies that heat dissipation has to be considered very carefully at the transition. There is usually no imaginary component in the impedance of a superconductor at low frequencies. But an AC signal inevitably produces a time varying temperature in the superconductor when its resistance is finite close to T$_c$. This leads to an imaginary component in the impedance which depends on thermal circuit describing the connection of the superconductor with its environment\cite{EnssBook}. While this has been known since long and is routinely used to characterize TES\cite{Jones:53,EnssBook,doi:10.1063/1.1889427,doi:10.1063/1.1559000}, it has not been remarked nor used (to our knowledge) in studies of the resistive transition in superconducting compounds\cite{Varlamov,RevModPhys.90.015009}. 

Techniques to study the resistive transition in a superconductor are numerous, but are mostly restricted to electrical measurements\cite{Varlamov,RevModPhys.90.015009}. There are however relevant open questions, which require an additional tool providing access to thermal properties. Close to the transition, Cooper pairs coexist with normal quasiparticles in an out-of-equilibrium quantum liquid whose thermal behavior is still largely unknown\cite{Schmid1966,Schmid1975207,PhysRevLett.28.1363,Chen2014}. Usual specific heat measurements are made with zero applied current through the sample and an external heater and thermometer. When applying a current through the sample close to T$_c$, the sample dissipates heat itself and it can be quite difficult to measure the temperature by an external thermometer. But at the resistive transition, the sample is itself heater and thermometer. We hereafter show that impedance measurements can be used as a thermal probe of the superconducting transition.

We make detailed real and imaginary impedance measurements of a 2H-NbSe$_2$ single crystal. We study a large imaginary component in the impedance at the transition and measure the temperature dependence of the imaginary component as a function of frequency, magnetic field and applied current. We take expressions for heat and current flow developed for TES and use these successfully to reproduce our result. We show that He exchange gas modifies the thermal connection and characterize it using the imaginary component. We also obtain the temperature dependence of the specific heat close to the transition. For small applied currents, we find a result compatible with macroscopic specific heat experiments, with a peak at T$_c$ of order of the electronic contribution to the specific heat. Under magnetic fields and with large currents, we find an increased peak, suggesting that vortex motion and out-of-equilibrium quasiparticles influence the heat balance.

\section{Experiment and methods}
 
\begin{figure}
	\includegraphics[width=0.45\textwidth]{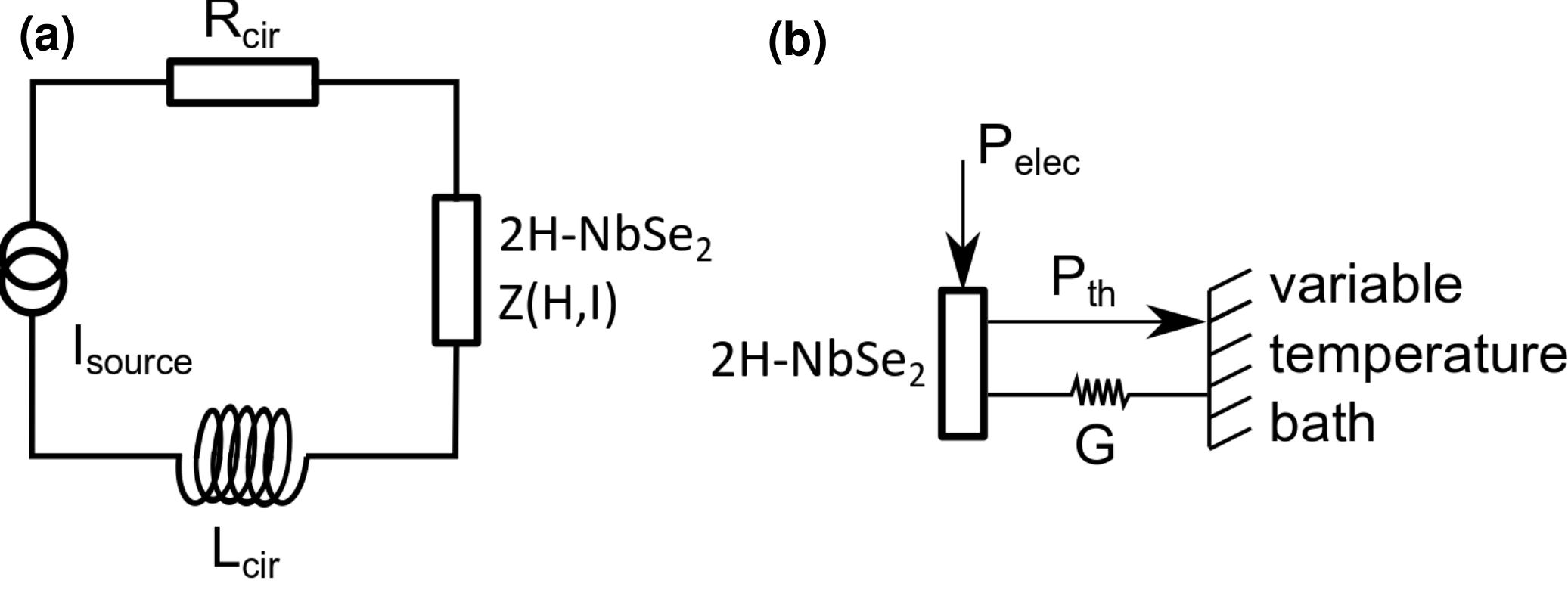}
	\caption{We show in (a) the electrical scheme of our setup. A current source (I$_{source}$) is connected to the 2H-NbSe$_2$ sample with impedance $Z$ through wiring which has a finite resistance $R_{circ}$ and an inductance $L_{circ}$. $I(t)$ follows eqn.1 in the text. In (b) we show the thermal scheme. An electrical power $P_{elec}$ is introduced as an electrical current in the 2H-NbSe$_2$ sample (box). This produces heat that flows to the thermal bath (P$_{th}$) through the thermal connection $G$. $T(t)$ follows eqn.2 and is connected to $I(t)$ through $P_{elec}=RI(t)^2$.}
	\label{FigScheme}
	\end{figure}

2H-NbSe$_2$ crystals were grown by iodine vapor transport, the crystals display the usual 2H-NbSe$_2$ properties: a feature at the charge density wave transition in the resistivity, a T$_c$ of about 7.2 K and a residual resistance ratio above 30, which gives an electronic mean free path above 100 nm (well above the superconducting coherence length of $\approx$ 10 nm)\cite{doi:10.1143/JPSJ.51.219}. We thinned down a 2H-NbSe$_2$ single crystal by exfoliation to about 14 $\mu$m thickness and glued the sample onto a Copper sample holder. The sample was 1.6 mm long and 0.8 mm wide and we used a thin layer of Kapton to separate it electrically from the sample holder. We used carefully thermalized twisted pair wires on a pumped helium bath cryostat equipped with a superconducting coil and a temperature controller. We measured the impedance using a lock-in amplifier and carefully monitored the phase shift and its temperature dependence. The highest applied currents correspond to a current density of about $2\times 10^6$ $A/m^2$, which is five orders of magnitude below the depairing current density $J_d\approx 10^{11}$ $A/m^2$\cite{PhysRevB.88.064518}. The power used is of about $50$ $\mu$W at the largest currents.

We describe schematically our electrical and thermal circuits in Fig\,\ref{FigScheme}. The current flows through the resistance and the wires (with resistance $R_{cir}$ and inductance $L_{cir}$, Fig\,\ref{FigScheme}(a)). The inductance $L_{cir}$ is of the order of a few $\mu$H and $R_{cir}$ is much smaller than the samples' resistance. The voltage induced by the oscillatory current is given by $L\frac{d I}{d t}$ ($L$ is the sum of the inductance of the wiring and of the kinetic inductance $L_K$ of the sample). This equals the voltage drop at the sample, $ZI$ plus the voltage drop at the resistances of the circuit\cite{EnssBook,Ullom_2015,Redfern2002,Lindeman2004,Taralli2010a}:

\begin{equation}
-L\frac{dI}{dt}=(R_{Source}+R_{cir})I+ZI,
\end{equation}

A time dependent current in the sample produces a time dependent temperature in the sample too. The Joule power $P_{elec}=RI^2$ released in the sample, minus the power leaking through the thermal link to the bath, $P_{th}$ (Fig.\,\ref{FigScheme}(b)), is equal to the power that the sample absorbs, which is the heat capacity $C$ of the sample times $\frac{dT}{dt}$\cite{EnssBook,Ullom_2015,Redfern2002,Lindeman2004,Taralli2010a}:

\begin{equation}
C\frac{dT}{dt}=R I^2-P_{th},
\end{equation}

\begin{figure*}
	\includegraphics[width=0.9\textwidth]{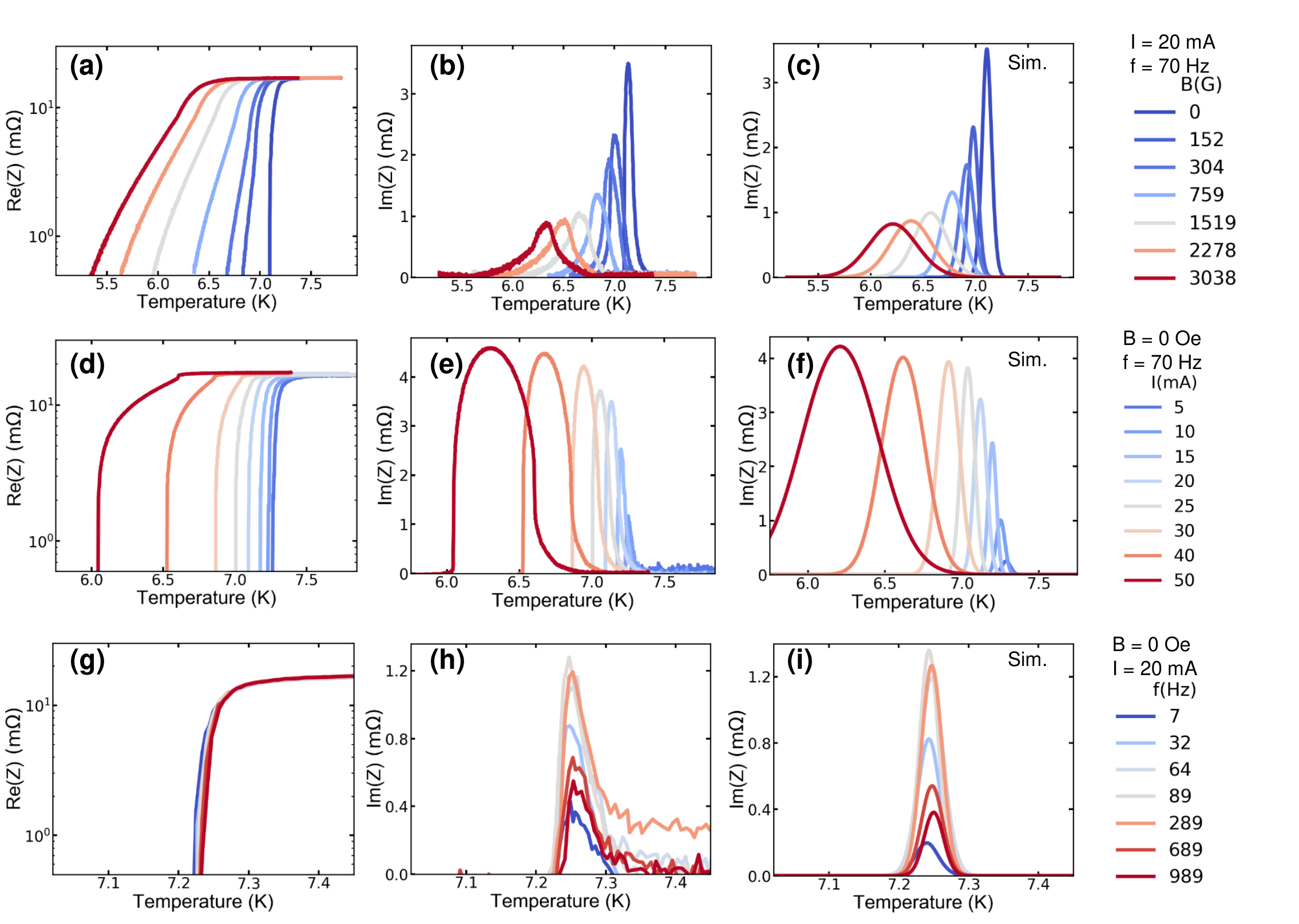}
	\caption{In (a) we show the real impedance vs temperature for different magnetic fields (lines from blue to red: 0 Oe, 152 Oe, 304 Oe, 759 Oe, 1519 Oe, 2278 Oe, and 3038 Oe). In (b) we show the imaginary impedance vs temperature for the same magnetic fields. Data in (a,b) are taken at 20 mA and a frequency of 70 Hz.  In (d) we show the real impedance for different values of the current (from blue to red: 5 mA, 10 mA, 15 mA, 20 mA, 25 mA, 30 mA, 40 mA and 50 mA). In (e) we show the imaginary impedance vs temperature for the same current values. Data in (d,e) are taken at zero magnetic field and a frequency of 70 Hz. In (g) we show the real impedance vs temperature for different frequencies (from blue to red: 7 Hz, 32 Hz, 64 Hz, 89 Hz, 289 Hz, 689 Hz, 989 Hz). In (h) we show the imaginary impedance for the same frequencies. Data in (g,h) are taken at zero magnetic field and a current of 20 mA. In (c,f,i) we show the imaginary impedance using Eqn.\ref{Sol_Imag} and the approximations described in the text. For clarity, we show the color scale in each set of figures as bars on the right.}
	\label{FigMagField}
	\end{figure*}

We can introduce the parameters $\alpha=\frac{T}{R}\frac{\partial R}{\partial T}$ and $\beta = \frac{I}{R}\frac{\partial R}{\partial I}$, which are the logarithmic derivatives of $R=Re(Z)$ with temperature and current \cite{EnssBook,Ullom_2015,Redfern2002,Lindeman2004,Taralli2010a}, and perform a local linearization to write the impedance versus frequency $Z(\omega)$:

\begin{equation}
\begin{split}
Z(\omega)=i\omega L + R_{cir}+R(1+\beta)+\\
\frac{2+\beta}{1+i\omega\frac{CT}{GT-I^2R\alpha}}\frac{R^2I^2\alpha/(GT)}{1-\frac{I^2R\alpha}{GT}}.
\end{split}
\label{Solution}
\end{equation}

The reactance is the imaginary part of $Z$, $Im(Z)$: 
\begin{equation}
\begin{split}
Im(Z)(\omega)=\omega L -
\omega\frac{CT(2+\beta)}{1+\left|\omega\frac{CT}{GT-I^2R\alpha}\right|^2}\frac{R^2I^2\alpha/(GT)^2}{\left(1-\frac{I^2R\alpha}{GT}\right)^2}.
\end{split}
\label{Sol_Imag}
\end{equation}

The parameters $T$, $R$, $I$ and $\alpha$ are measured, whereas $\beta$, $G$ and $C$ are determined by comparing the measured $Im(Z)(\omega)$ to this expression. The reactance has a maximum at a frequency of $\frac{GT-I^2R\alpha}{CT}=\frac{G}{C}-\frac{I^2dR}{CdT}$, which is the difference between the inverse of the thermal time constant of the system $\frac{G}{C}$ and the ratio between the differential Joule power $P_{elec}=I^2dR$ and the differential power admitted by the sample ${CdT}$.

\begin{figure}
	\includegraphics[width=0.45\textwidth]{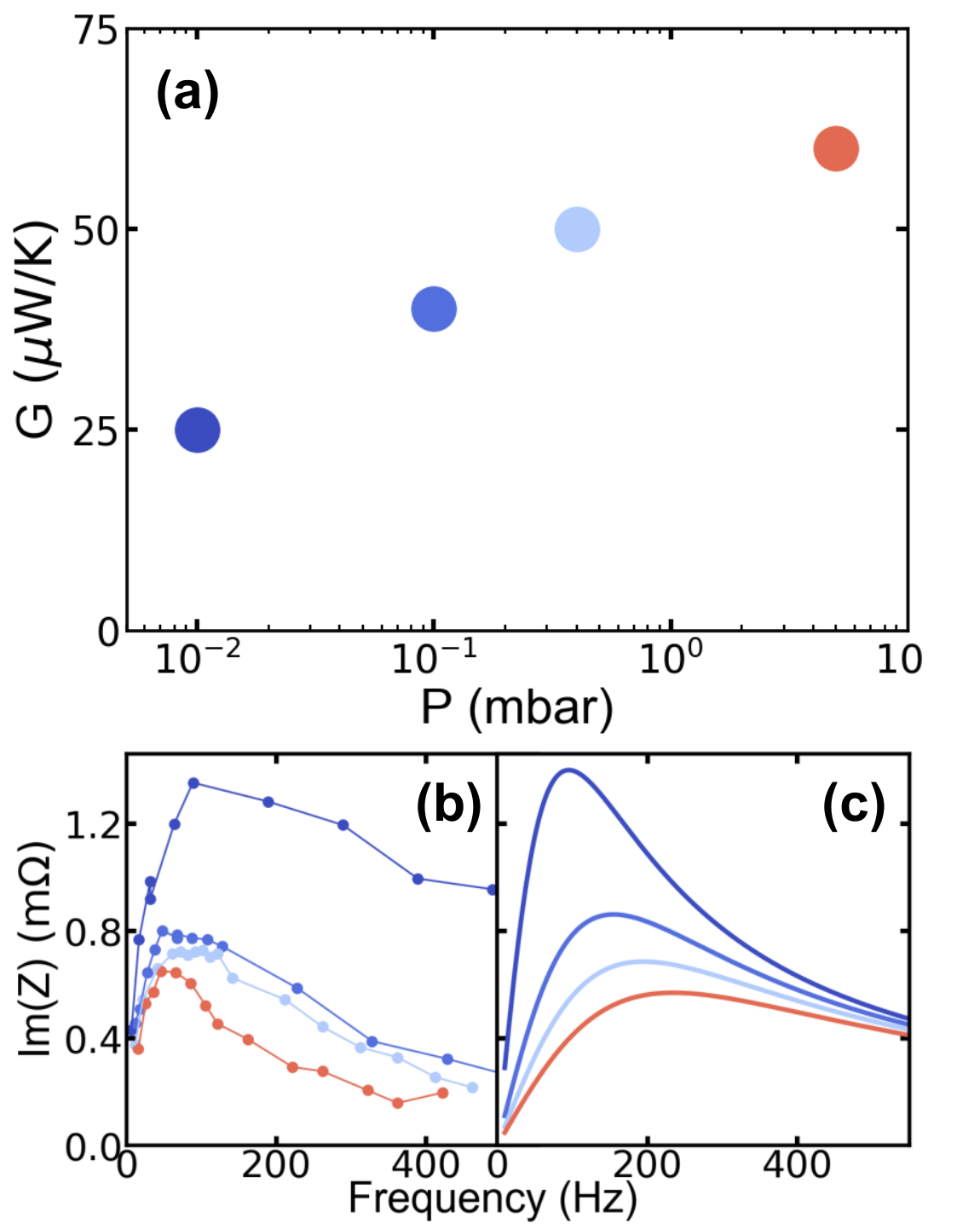}
	\caption{(a) We show the thermal conductance vs exchange gas pressure curve deduced from Eqn.\ref{Sol_Imag}. In inset (b) we show the imaginary impedance as a function at frequency at about 7.2K. The exchange gas pressure goes from 0.01 mbar to 5 mbar. The lines color in (b) and (c) correspond to the values shown in the insets (a).  In inset (c) we show the result of the calculation described in the text that uses Eqn.\ref{Sol_Imag}.}
	\label{FigPressure}
	\end{figure}

\section{Results and discussion}

In Fig.\,\ref{FigMagField}(a,b) we show the real and imaginary impedance. At zero magnetic field (blue curve) we observe a really sharp transition in the real impedance. We observe, at the same time, a strong and sharp peak in the imaginary impedance. The magnetic field reduces the superconducting critical temperature and broadens the resitive transition (Fig.\,\ref{FigMagField}(a)). The decay of the real impedance with temperature is exponential for low values of the impedance. The imaginary impedance decreases and is broader than at zero field.

In Fig.\,\ref{FigMagField}(d,e) we show the effect of increasing the current at zero field. We observe that the transition in the real impedance broadens. The peak in the imaginary impedance, however, strongly increases in size, reaching approximately 30\% of the value of the real impedance. The temperature range with a high real impedance results in a broad peak in the imaginary impedance. The imaginary impedance vanishes exponentially at low temperatures, following the real impedance.

In Fig.\,\ref{FigMagField}(g,h), we show the effect of modifying the frequency at a fixed current and for zero magnetic field. The real impedance remains unaffected, but the imaginary impedance first increases up to about 70 Hz and then decreases when approaching 1 kHz. Thus, there is a frequency range, of the order of a few tens of Hz, where the imaginary impedance is largest. In the Appendix A we provide the imaginary part of the impedance at a fixed point at the transition as a function of frequency.

Let us start by discussing the usual electrodynamic frequency response of superconductors. The kinetic inductance $L_K$ provides a finite reactance in the superconducting phase\cite{Schmidt2009a}. We can estimate the kinetic inductance $L_K$ using $L_K=(4\pi \lambda^2)/d=3.6 \times 10^{-8}$ H with $\lambda=200$ nm the penetration depth of 2H-NbSe$_2$\cite{Takita1985} and $d$ the thickness of the sample. This provides a contribution to the reactance three orders of magnitude below our observations of $3.6\times 10^{-3}$  $m\Omega$ at 100 Hz. Furthermore, the maximum of the kinetic impedance occurs at frequencies in the GHz range, whereas we work here at frequencies well below a kHz and the maximum in the reactance occurs at merely 70 Hz (Fig.\,\ref{FigMagField}(h,i)). Thus, the kinetic inductance does not explain the observed behavior.

We can also calculate the thermal diffusion length scale $L_{thermal}=\frac{\kappa}{C\tau\rho}$ (where $\kappa$ is the samples' thermal conductivity, $\tau$ the time scale for the variations in the current and $\rho$ the density) and see that it is much larger than the sample size (about 3 mm at 70 Hz and a cm at 1 Hz). Thus, there are no temperature induced gradients induced within the sample by the applied current.

We now consider the coupling between electronic and heat transport, using the model described previously and the Eqn.\ref{Sol_Imag}. We start with a temperature independent heat capacity of $C\approx 1.7 \times 10^{-7}$  J/K, a temperature independent thermal conductance of $G\approx 1.5 \times 10^{-5}$  J/K, we assume a temperature independent parameter $\beta\approx 1.5$. In most of the temperature and current range we are considering, the I-V characteristics of NbSe$_2$ is in the vortex liquid or flux-flow regimes, which leads to a smooth, non-exponential behavior. Furthermore, we start by taking for $\frac{dRe(Z)}{dT}$ vs temperature a Gaussian form centered at the midpoint of the transition. We replace R in Eqn.\ref{Sol_Imag} with the measured impedance $Re(Z)$. We can then calculate the reactance using Eqn.\ref{Sol_Imag}. We obtain the results shown in Fig.\,\ref{FigMagField}(c,f,i). The order of magnitude of the imaginary impedance and of its temperature dependence is similar than the ones observed in the experiment Fig.\,\ref{FigMagField}(b,e,h).

We thus see that the broadening of the transition obtained as a function of the magnetic field in the real impedance $Re(Z)$ (Fig.\,\ref{FigMagField}(a)) results in a strong decrease of the imaginary impedance $Im(Z)$ (Fig.\,\ref{FigMagField}(b)) at the transition. On the other hand, the increase of the current I, with the concomitant broadening of the transition in the real impedance $Re(Z)$, produces the opposite in the imaginary impedance $Im(Z)$. $Im(Z)$ (Fig.\,\ref{FigMagField}(e)) strongly increases with current. Finally, as a function of frequency, we find that the imaginary impedance $Im(Z)$ shows a peak at about 70 Hz and that it vanishes for low and high frequencies, as also discussed in more detail in Appendix A. Thus, the order of magnitude of the effect is very well captured by the thermal model, in spite of the approximations used.

Let us now discuss the dependence as a function of the exchange gas pressure, shown in Fig.\,\ref{FigPressure}. Usually, exchange gas improves the coupling of the sample to its thermal environment. It thus primarily increases the thermal conductivity $G$. Interestingly, the imaginary component $Im(Z)(\omega)$ provides a rather accurate account of the exchange gas present in the experiment. To see this, we have measured $Im(Z)(\omega)$ as a function of the frequency for different exchange gas residual pressures. The frequency dependence for vanishing exchange gas pressure follows the one observed in Fig.\,\ref{FigMagField}(g,h) and discussed in Appendix A. We use our model to obtain $G$ for each exchange gas pressure.

Fig.\,\ref{FigPressure}(a) displays the obtained thermal conductance $G$ vs the He exchange gas pressure. In Fig.\,\ref{FigPressure}(b) we show $Im(Z)(\omega)$ for different exchange gas pressures. We observe that $Im(Z)(\omega)$ strongly decreases with increasing exchange gas pressure. It is thus a good measurement of the residual exchange gas present in the experiment. The frequency dependence $Im(Z)(\omega)$ calculated using our model, changing the value of $G$, is shown in Fig.\,\ref{FigPressure}(c). We see that the calculations provide a good account of the observed overall decrease in $Im(Z)(\omega)$. There are some aspects, like the dependence of the position of the maximum with frequency, which are not precisely captured by the model. Nevertheless, it is quite remarkable that the values of $G$ as well as its dependence on the residual exchange gas pressure are in agreement with direct measurements of the thermal conductance of He exchange gas\cite{ganta2011optical}. Thus, the improved heat transport with the thermal bath through convection by the exchange gas clearly leads to a thermal behavior of the sample which is less influenced by small oscillations in temperature.

\begin{figure}
	\includegraphics[width=0.47\textwidth]{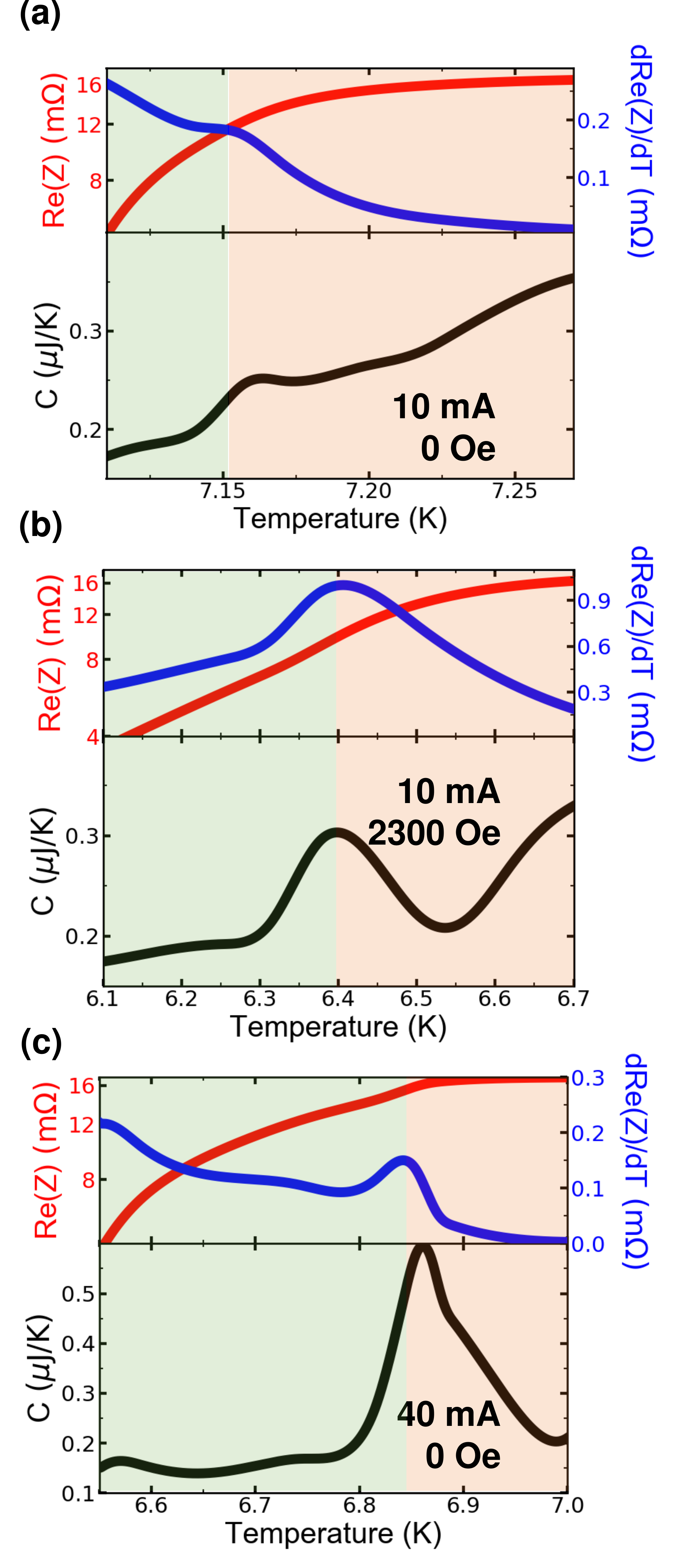}
	\caption{In the top panels of each figure we show the temperature dependence of the resistance (left y-axis, red color) and of $\frac{dR}{dT}$ (right y-axis, blue color). In the bottom panel we show the value of $C$ obtained as discussed in the text (black line). Notice that the temperature range shown in each figure corresponds to the range where the reactance is finite. In (a) we show results obtained at 10 mA and zero magnetic field (blue curve in Fig.\ref{FigMagField}(a,b)), in (b) at 10 mA and 2300 Oe (orange curve in Fig.\ref{FigMagField}(a,b)) and in (c) results at 40 mA and zero magnetic field (orange curve in in Fig.\ref{FigMagField}(d,e)). Green shaded temperature range corresponds to the temperature region corresponds to the temperature region where at least part of the sample is superconducting and rosa shading to the normal region dominated by fluctuation and dissipation.}
	\label{FigSpecificHeat}
	\end{figure}

We also see that there are slight differences between the calculated (Fig.\,\ref{FigMagField}(c,f,i)) and measured (Fig.\,\ref{FigMagField}(b,e,h)) imaginary impedances as a function of the magnetic field and current, for vanishing residual exchange gas. Assuming that these differences are just due to the temperature variation of $C$ in Eqn.\ref{Sol_Imag}, we calculate $C$ as a function of temperature. For this, we use the temperature dependence of $Re(Z)$ obtained from the experiment to numerically calculate $\frac{dRe(Z)}{dT}$. We compute $Im(Z)$ starting with an uniform Ansatz curve for $C(T)$ and vary it numerically until we obtain the measured temperature dependence of $Im(Z)$. In Fig.\,\ref{FigSpecificHeat} we show the result for a three characteristic situations.

It is useful to discuss the obtained $C(T)$ together with the temperature dependence of the resistance and of its derivative. At zero magnetic field and with a small current (Fig.\,\ref{FigSpecificHeat}(a)), we observe that the resistance drops continuously with decreasing temperature. But $\frac{dRe(Z)}{dT}$ does not increase smoothly until it diverges at the transition, it shows a peak at approximately 7.15 K. This leads to a small peak in $C(T)$. The value we find for $C$ is comparable to the estimated heat capacity of our sample. Its temperature dependence is similar as the one observed in the heat capacity of 2H-NbSe$_2$ using macroscopic measurements. The heat capacity of the sample increases by the same amount in the temperature range shown in Fig.\,\ref{FigSpecificHeat}(a) \cite{Huang2007a,PhysRevLett.90.117003}. The peak in $C$ can be related to the peak in $C$ at the superconducting transition and is of roughly the same order.

When applying a magnetic field, the temperature range with a finite $\frac{dRe(Z)}{dT}$ inside the superconducting phase becomes considerably larger  (Fig.\,\ref{FigSpecificHeat}(b)). The peak in $C$ also becomes larger.

When applying a current at zero magnetic field, (Fig.\,\ref{FigSpecificHeat}(c)), the peak in $C$ becomes even larger and there is a small but finite $C$ over a substantial temperature range.

This temperature independent $C$ well within the superconducting phase in presence of a large current (at low temperatures in Fig.\,\ref{FigSpecificHeat}(c)) is quite remarkable. It corresponds roughly to the temperature range where $Im(Z)$ shows a broad maximum. Thus, there is a mechanism for heat production that develops in presence of large currents well within the superconducting phase. This is related to vortex motion and the associated generation of quasiparticles.

First we should realize that in Fig.\,\ref{FigSpecificHeat}(c) the temperature is close to T$_c$ and the current above the critical current for the onset of vortex motion. In this range, the transition is very broad and parameters such as $\alpha$ and $\beta$ have a small and smooth temperature dependence. When applying a current, vortices enter the sample. Vortices are pinned at defects and are mobile in between pinning centers in presence of a current\cite{Blatter1994}. In this temperature range, and in presence of such large currents, vortices are mobile. During vortex motion, the Lorentz force is compensated by a drag force which is dissipative\cite{PhysRev.139.A1163,MAKI1971124,PhysRev.140.A1197,PhysRevB.6.110,PhysRevLett.20.735,Schmid1966}. Moving vortex cores requires transforming normal quasiparticles into Cooper pairs and produces out of equilibrium quasiparticles along their path\cite{PhysRevB.97.094510}. At large driving currents, vortices move at very high velocities, even higher than the speed of sound\cite{Embon2017,PhysRevB.92.024513,PhysRevApplied.11.054064}. They can be unstable at high driving velocities, leading to additional quasiparticles\cite{Embon2017,Estelles-Duart2018,Jing_2018,PhysRevApplied.11.054005,PhysRevLett.97.067003,PhysRevLett.72.752}. The observed increase in the imaginary impedance under magnetic fields shows the contribution from fluctuating vortices to the superconducting transition.

The value of $C$ we find is of order of the electronic contribution to the specific heat at zero current. Thus, the amount of excitations created by the current remains approximately constant in the temperature range when the real component of the impedance is finite, in spite of a strongly decreasing real impedance. The real impedance measures the voltag induced by current flow, which results from vortex motion between pinning centers. The imaginary impedance measures instead the heat created in this process.

\section{Conclusion and outlook}

In conclusion, we observe a strong mutual influence between heat and electronic transport in the superconducting transition of 2H-NbSe$_2$. The frequency for the appearance of the coupling is in the tens of Hz regime. We use a model which traces back the behavior of the reactance from the resistance, taking into account heating effects.

Our results suggest that both real and imaginary impedance measurements are fundamental to determine the effect of temperature on a sharp superconducting transition. As we discuss in more detail in Appendix B, the measurement of the imaginary impedance should be very useful to characterize superconducting transitions, because it highlights overheating effects or the presence of residual exchange gas. More than that, as we show with the data as a function of current, it also provides precise information for the heat capacity of the sample. For example, a fundamental aspect in two-dimensional systems is the presence of out of equilibrium dissipation and coherence at the same time, as a consequence of a continous, Kosterlitz-Thouless type of transition into the normal state\cite{Chen2014}. Very recent measurements suggest thermally driven vortex blockade in ultra thin devices of 2H-NbSe$_2$\cite{benyamini2019blockade}. Measurements of the critical current in 2H-NbSe$_2$ contacted with graphene show strongly reduced values with respect to a metallic electrode, suggesting that electron flow in graphene generates heat that is transferred to 2H-NbSe$_2$\cite{PhysRevB.98.035422}. These measurements consider only the real impedance, which just shows electronic transport. The imaginary impedance should be much larger in thin films in the limit of small currents than we observe here and can serve as a new method to characterize the thermal behavior in these and similar systems.

Our results also show the highly non-linear effect of the exchange gas (Fig.\,\ref{FigPressure}). The measured dependence of the thermal conductance $G$ due to mass flow (convection) of gas in vacuum as a function of the pressure shows a similar increase as we observe here\cite{Funke2015a,Ganta2011a}. The order of magnitude of $G$ corresponds to a distance of the order of a cm, which is comparable to the size of our set-up. Recent measurements of the dissipation in quantum systems have been made by connecting the temperature sensor to the sample through exchange gas\cite{Halbertal2016,Halbertal2017}. The measurement of real and imaginary components of the resistive superconducting transition can be used to independently characterize this link.

One might expect at first sight that thermal effects are just a consequence of having a sharp transition. Our work shows that the broadening of the transition by the application of current does not lead to a vanishing imaginary component. We reveal a strong increase of the imaginary impedance when applying large currents. This increase is due to an additional contribution to the specific heat from the quasiparticles generated during vortex motion.

\section{Acknowledgments}
This work was supported by the Spanish Research State Agency (FIS2017-84330-R, MDM-2014-0377, MAT2017-89993-R and MDM-2015-0538), by the Comunidad de Madrid through program NANOFRONTMAG-CM (S2013/MIT-2850), by the European Research Council PNICTEYES grant agreement 679080 and by EU program Cost CA16218 (Nanocohybri). We also acknowledge SEGAINVEX at UAM. We thank R. \'Alvarez Montoya for technical support, A. Garc\'ia and D. Caldevilla for support at the beginning of the project. We also thank enlightening discussions with Andrey Varlamov, A.I. Buzdin and J.C. Cuevas.

\section{Appendix A: Imaginary impedance vs frequency}

\begin{figure}
	\includegraphics[width=0.45\textwidth]{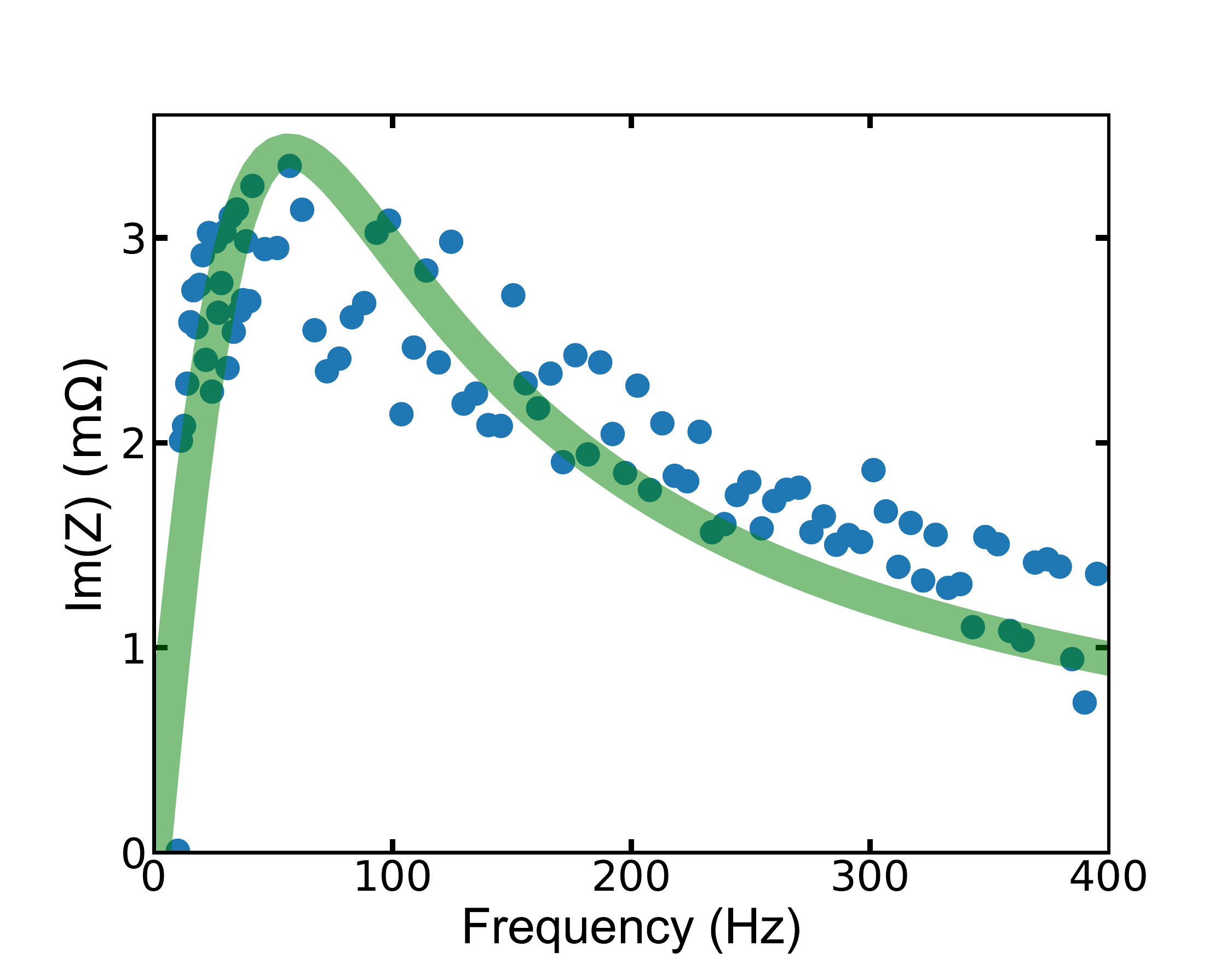}
	\caption{We show (blue points) the reactance vs frequency, the fits to the model described in the text is shown in green. The measurement temperature is 7.2 K and the exchange gas residual pressure is below 0.01 mbar. }
	\label{FigFrequency}
	\end{figure}

We present here measurement of the impedance at a fixed temperature while varying the frequency. To obtain the green line in Fig.\,\ref{FigFrequency} we use $C=4.09\times 10^{-8}$ J/K, $G=1.5 \times 10^{-5}$ W/K, $\alpha=65$, $\beta=15$ which are the values obtained by fitting the imaginary part of the impedance with Eqn.\ref{Sol_Imag}. The heat capacity $C$ of our sample can be estimated through the sample size (1.6 mm$\times$ 0.8 mm $\times$ 14 $\mu$m) and the molar heat capacity of 2H-NbSe$_2$ $400$ mJ/(mol K) \cite{Huang2007a}, we obtain $C\approx 1.7 \times 10^{-7}$  J/K. With the value of $G$ used we can estimate the $dT$ produced by the Joule power in the sample and obtain about 20 mK. Using $\alpha$, we can again estimate $dT$ and obtain approximately the same value. The obtained $\alpha$ and $\beta$ are compatible with usual values at the transition in 2H-NbSe$_2$. The agreement between calculations and experiment (Fig.\,\ref{FigFrequency}) and the values obtained for the different parameters show that the linearized equations account well for the behavior obtained in a large part of the superconducting transition.

\section{Appendix B: Comments on the measurement of the resistive transition}

Our result implies that the resistive transition cannot be understood solely on the basis of measurements of the real impedance. The resistive transition of superconductors has been studied in depth in the limit of vanishing current, or when heat dissipation in the sample can be neglected\cite{RevModPhys.90.015009,PhysRevB.8.4164,Kremen2018,Varlamov}. The discussion has focused on the influence of fluctuations on the conductivity. When approaching the transition from higher temperatures, fluctuations modify the conductance gradually from the normal state value until it diverges at some point. The temperature range for influence of fluctuations in thermodynamic properties is approximately given by $\frac{T-T_c}{Tc}>G_i$, which in 2H-NbSe$_2$ is practically negligible\cite{Larkin2008}. However, in the conductivity, fluctuations appear much earlier due to nonlinear effects, at about $\frac{T-T_c}{Tc}>\sqrt{G_i}$, which leads to a temperature range that can cover a few tens of mK\cite{Larkin2001}. Indeed, at the smallest currents we observe that the resistance starts to drop a few tens of mK before the actual transition. Different contributions might modify the conductivity at zero magnetic field and zero frequency around the superconducting transition in a superconductor. First, strongly time dependent fluctuations of the superconducting order parameter, that lead, averaged over time, to bubbles with higher conductance due to time fluctuating preformed Cooper pairs and is termed the Aslamazov-Larkin contribution\cite{RevModPhys.90.015009}. Second, the normal state density of states might show a dip already above $T_c$\cite{RevModPhys.90.015009}. Third, the Maki-Thompson contribution, due to the formation of Cooper pairs at self-interfering trajectories caused by scattering at impurities\cite{10.1143/PTP.39.897,PhysRevB.1.327,Carballeira2000}. In the clean limit (as we mention above, $\ell>>\xi$ in 2H-NbSe$_2$), only the Aslamazov-Larkin contribution is relevant and leads to the observed decrease in the resistance above T$_c$. Our results show that there is an imaginary impedance which remained unnoticed in the fluctuation range. The result at low currents and zero magnetic field seems to follow well the specific heat of the sample, suggesting that these mechanisms have a minor contribution to the heat balance in the sample. However, this changes when applying a magnetic field or a current, as we discuss in the main text.

On more experimental grounds, we should note that the imaginary impedance appears at low frequencies and thus in transport experiments that are not made exactly in DC conditions. All kinds of electronic measurements imply a change of the parameters with time, either to remove thermoelectric voltages in a usual four-wire measurement \cite{Pope2001a,QuantumDesign1999} or simply to vary the temperature in regular steps. Through the power used to measure, there is a connection between the resistance and temperature, which induces a reactance when the resistance is strongly temperature dependent.

\bibliographystyle{apsrev4-1-titles}

\begin{thebibliography}{51}%
\makeatletter
\providecommand \@ifxundefined [1]{%
 \@ifx{#1\undefined}
}%
\providecommand \@ifnum [1]{%
 \ifnum #1\expandafter \@firstoftwo
 \else \expandafter \@secondoftwo
 \fi
}%
\providecommand \@ifx [1]{%
 \ifx #1\expandafter \@firstoftwo
 \else \expandafter \@secondoftwo
 \fi
}%
\providecommand \natexlab [1]{#1}%
\providecommand \enquote  [1]{``#1''}%
\providecommand \bibnamefont  [1]{#1}%
\providecommand \bibfnamefont [1]{#1}%
\providecommand \citenamefont [1]{#1}%
\providecommand \href@noop [0]{\@secondoftwo}%
\providecommand \href [0]{\begingroup \@sanitize@url \@href}%
\providecommand \@href[1]{\@@startlink{#1}\@@href}%
\providecommand \@@href[1]{\endgroup#1\@@endlink}%
\providecommand \@sanitize@url [0]{\catcode `\\12\catcode `\$12\catcode
  `\&12\catcode `\#12\catcode `\^12\catcode `\_12\catcode `\%12\relax}%
\providecommand \@@startlink[1]{}%
\providecommand \@@endlink[0]{}%
\providecommand \url  [0]{\begingroup\@sanitize@url \@url }%
\providecommand \@url [1]{\endgroup\@href {#1}{\urlprefix }}%
\providecommand \urlprefix  [0]{URL }%
\providecommand \Eprint [0]{\href }%
\providecommand \doibase [0]{http://doi.org/}%
\providecommand \selectlanguage [0]{\@gobble}%
\providecommand \bibinfo  [0]{\@secondoftwo}%
\providecommand \bibfield  [0]{\@secondoftwo}%
\providecommand \translation [1]{[#1]}%
\providecommand \BibitemOpen [0]{}%
\providecommand \bibitemStop [0]{}%
\providecommand \bibitemNoStop [0]{.\EOS\space}%
\providecommand \EOS [0]{\spacefactor3000\relax}%
\providecommand \BibitemShut  [1]{\csname bibitem#1\endcsname}%
\let\auto@bib@innerbib\@empty
\bibitem [{\citenamefont {Irwin}\ and\ \citenamefont
  {G.C.~Hilton}(2005)}]{EnssBook}%
  \BibitemOpen
  \bibfield  {author} {\bibinfo {author} {\bibfnamefont {K.}~\bibnamefont
  {Irwin}}\ and\ \bibinfo {author} {\bibfnamefont {E.~b. C.~E.}\ \bibnamefont
  {G.C.~Hilton}},\ }\href {https://link.springer.com/book/10.1007/b12169#toc}
  {\emph {\bibinfo {title} {"Transition edge sensors" in "Cryogenic particle
  detection"}}},\ \bibinfo {edition} {1st}\ ed.,\ \bibinfo {series} {Topics
  Appl. Phys.}, Vol.~\bibinfo {volume} {99}\ (\bibinfo  {publisher} {Springer -
  Verlag},\ \bibinfo {address} {Berlin-Heidelberg-New York},\ \bibinfo {year}
  {2005})\BibitemShut {NoStop}%
\bibitem [{\citenamefont {Ullom}\ and\ \citenamefont
  {Bennett}(2015)}]{Ullom_2015}%
  \BibitemOpen
  \bibfield  {author} {\bibinfo {author} {\bibfnamefont {J.~N.}\ \bibnamefont
  {Ullom}}\ and\ \bibinfo {author} {\bibfnamefont {D.~A.}\ \bibnamefont
  {Bennett}},\ }\bibfield  {title} {\emph {\bibinfo {title} {{Review} of
  superconducting transition-edge sensors for {X}-ray and {$\gamma$}-ray
  spectroscopy}},\ }\href {\doibase 10.1088/0953-2048/28/8/084003} {\bibfield
  {journal} {\bibinfo  {journal} {Superconductor Science and Technology}\
  }\textbf {\bibinfo {volume} {28}},\ \bibinfo {pages} {084003} (\bibinfo
  {year} {2015})}\BibitemShut {NoStop}%
\bibitem [{\citenamefont {Redfern}\ \emph {et~al.}(2002)\citenamefont
  {Redfern}, \citenamefont {Nicolosi}, \citenamefont {H{\"o}hne}, \citenamefont
  {Weiland}, \citenamefont {Simmnacher},\ and\ \citenamefont
  {Hollerich}}]{Redfern2002}%
  \BibitemOpen
  \bibfield  {author} {\bibinfo {author} {\bibfnamefont {D.}~\bibnamefont
  {Redfern}}, \bibinfo {author} {\bibfnamefont {J.}~\bibnamefont {Nicolosi}},
  \bibinfo {author} {\bibfnamefont {J.}~\bibnamefont {H{\"o}hne}}, \bibinfo
  {author} {\bibfnamefont {R.}~\bibnamefont {Weiland}}, \bibinfo {author}
  {\bibfnamefont {B.}~\bibnamefont {Simmnacher}}, \ and\ \bibinfo {author}
  {\bibfnamefont {C.}~\bibnamefont {Hollerich}},\ }\bibfield  {title} {\emph
  {\bibinfo {title} {The microcalorimeter for industrial applications}},\
  }\href {\doibase 10.6028/jres.107.050} {\bibfield  {journal} {\bibinfo
  {journal} {Journal of research of the National Institute of Standards and
  Technology}\ }\textbf {\bibinfo {volume} {107}},\ \bibinfo {pages} {621}
  (\bibinfo {year} {2002})}\BibitemShut {NoStop}%
\bibitem [{\citenamefont {Jones}(1953)}]{Jones:53}%
  \BibitemOpen
  \bibfield  {author} {\bibinfo {author} {\bibfnamefont {R.~C.}\ \bibnamefont
  {Jones}},\ }\bibfield  {title} {\emph {\bibinfo {title} {The general theory
  of bolometer performance}},\ }\href {\doibase 10.1364/JOSA.43.000001}
  {\bibfield  {journal} {\bibinfo  {journal} {J. Opt. Soc. Am.}\ }\textbf
  {\bibinfo {volume} {43}},\ \bibinfo {pages} {1} (\bibinfo {year}
  {1953})}\BibitemShut {NoStop}%
\bibitem [{\citenamefont {Vaillancourt}(2005)}]{doi:10.1063/1.1889427}%
  \BibitemOpen
  \bibfield  {author} {\bibinfo {author} {\bibfnamefont {J.~E.}\ \bibnamefont
  {Vaillancourt}},\ }\bibfield  {title} {\emph {\bibinfo {title} {Complex
  impedance as a diagnostic tool for characterizing thermal detectors}},\
  }\href {https://doi.org/10.1063/1.1889427} {\bibfield  {journal} {\bibinfo
  {journal} {Review of Scientific Instruments}\ }\textbf {\bibinfo {volume}
  {76}},\ \bibinfo {pages} {043107} (\bibinfo {year} {2005})}\BibitemShut
  {NoStop}%
\bibitem [{\citenamefont {Galeazzi}\ and\ \citenamefont
  {McCammon}(2003)}]{doi:10.1063/1.1559000}%
  \BibitemOpen
  \bibfield  {author} {\bibinfo {author} {\bibfnamefont {M.}~\bibnamefont
  {Galeazzi}}\ and\ \bibinfo {author} {\bibfnamefont {D.}~\bibnamefont
  {McCammon}},\ }\bibfield  {title} {\emph {\bibinfo {title} {Microcalorimeter
  and bolometer model}},\ }\href {\doibase 10.1063/1.1559000} {\bibfield
  {journal} {\bibinfo  {journal} {Journal of Applied Physics}\ }\textbf
  {\bibinfo {volume} {93}},\ \bibinfo {pages} {4856} (\bibinfo {year}
  {2003})}\BibitemShut {NoStop}%
\bibitem [{\citenamefont {Anatoly~Larkin}(2009)}]{Varlamov}%
  \BibitemOpen
  \bibfield  {author} {\bibinfo {author} {\bibfnamefont {A.~V.}\ \bibnamefont
  {Anatoly~Larkin}},\ }\href@noop {} {\emph {\bibinfo {title} {Theory of
  fluctuations in superconductors}}},\ \bibinfo {series} {International Series
  on Monographs in Physics}, Vol.\ \bibinfo {volume} {127}\ (\bibinfo
  {publisher} {Oxford University Press},\ \bibinfo {address} {Oxford, UK},\
  \bibinfo {year} {2009})\BibitemShut {NoStop}%
\bibitem [{\citenamefont {Varlamov}\ \emph {et~al.}(2018)\citenamefont
  {Varlamov}, \citenamefont {Galda},\ and\ \citenamefont
  {Glatz}}]{RevModPhys.90.015009}%
  \BibitemOpen
  \bibfield  {author} {\bibinfo {author} {\bibfnamefont {A.~A.}\ \bibnamefont
  {Varlamov}}, \bibinfo {author} {\bibfnamefont {A.}~\bibnamefont {Galda}}, \
  and\ \bibinfo {author} {\bibfnamefont {A.}~\bibnamefont {Glatz}},\ }\bibfield
   {title} {\emph {\bibinfo {title} {Fluctuation spectroscopy: From
  {R}ayleigh-{J}eans waves to {A}brikosov vortex clusters}},\ }\href {\doibase
  10.1103/RevModPhys.90.015009} {\bibfield  {journal} {\bibinfo  {journal}
  {Rev. Mod. Phys.}\ }\textbf {\bibinfo {volume} {90}},\ \bibinfo {pages}
  {015009} (\bibinfo {year} {2018})}\BibitemShut {NoStop}%
\bibitem [{\citenamefont {Schmid}(1966)}]{Schmid1966}%
  \BibitemOpen
  \bibfield  {author} {\bibinfo {author} {\bibfnamefont {A.}~\bibnamefont
  {Schmid}},\ }\bibfield  {title} {\emph {\bibinfo {title} {A time dependent
  {G}inzburg-{L}andau equation and its application to the problem of
  resistivity in the mixed state}},\ }\href {\doibase 10.1007/BF02422669}
  {\bibfield  {journal} {\bibinfo  {journal} {Physik der kondensierten
  Materie}\ }\textbf {\bibinfo {volume} {5}},\ \bibinfo {pages} {302} (\bibinfo
  {year} {1966})}\BibitemShut {NoStop}%
\bibitem [{\citenamefont {Schmid}\ and\ \citenamefont
  {Schon}(1975)}]{Schmid1975207}%
  \BibitemOpen
  \bibfield  {author} {\bibinfo {author} {\bibfnamefont {A.}~\bibnamefont
  {Schmid}}\ and\ \bibinfo {author} {\bibfnamefont {G.}~\bibnamefont {Schon}},\
  }\bibfield  {title} {\emph {\bibinfo {title} {Linearized kinetic equations
  and relaxation processes of a superconductor near {T$_c$}}},\ }\href
  {\doibase 10.1007/BF00115264} {\bibfield  {journal} {\bibinfo  {journal}
  {Journal of Low Temperature Physics}\ }\textbf {\bibinfo {volume} {20}},\
  \bibinfo {pages} {207} (\bibinfo {year} {1975})}\BibitemShut {NoStop}%
\bibitem [{\citenamefont {Clarke}(1972)}]{PhysRevLett.28.1363}%
  \BibitemOpen
  \bibfield  {author} {\bibinfo {author} {\bibfnamefont {J.}~\bibnamefont
  {Clarke}},\ }\bibfield  {title} {\emph {\bibinfo {title} {Experimental
  observation of pair-quasiparticle potential difference in nonequilibrium
  superconductors}},\ }\href {\doibase 10.1103/PhysRevLett.28.1363} {\bibfield
  {journal} {\bibinfo  {journal} {Phys. Rev. Lett.}\ }\textbf {\bibinfo
  {volume} {28}},\ \bibinfo {pages} {1363} (\bibinfo {year}
  {1972})}\BibitemShut {NoStop}%
\bibitem [{\citenamefont {Chen}\ \emph {et~al.}(2014)\citenamefont {Chen},
  \citenamefont {Lin}, \citenamefont {Snyder}, \citenamefont {Goldman},\ and\
  \citenamefont {Kamenev}}]{Chen2014}%
  \BibitemOpen
  \bibfield  {author} {\bibinfo {author} {\bibfnamefont {Y.}~\bibnamefont
  {Chen}}, \bibinfo {author} {\bibfnamefont {Y.-H.}\ \bibnamefont {Lin}},
  \bibinfo {author} {\bibfnamefont {S.~D.}\ \bibnamefont {Snyder}}, \bibinfo
  {author} {\bibfnamefont {A.~M.}\ \bibnamefont {Goldman}}, \ and\ \bibinfo
  {author} {\bibfnamefont {A.}~\bibnamefont {Kamenev}},\ }\bibfield  {title}
  {\emph {\bibinfo {title} {Dissipative superconducting state of
  non-equilibrium nanowires}},\ }\href {https://doi.org/10.1038/nphys3008}
  {\bibfield  {journal} {\bibinfo  {journal} {Nature Physics}\ }\textbf
  {\bibinfo {volume} {10}},\ \bibinfo {pages} {567 EP } (\bibinfo {year}
  {2014})}\BibitemShut {NoStop}%
\bibitem [{\citenamefont {Naito}\ and\ \citenamefont
  {Tanaka}(1982)}]{doi:10.1143/JPSJ.51.219}%
  \BibitemOpen
  \bibfield  {author} {\bibinfo {author} {\bibfnamefont {M.}~\bibnamefont
  {Naito}}\ and\ \bibinfo {author} {\bibfnamefont {S.}~\bibnamefont {Tanaka}},\
  }\bibfield  {title} {\emph {\bibinfo {title} {Electrical transport properties
  in {2H-NbS$_2$, -NbSe$_2$, -TaS$_2$ and -TaSe$_2$}}},\ }\href {\doibase
  10.1143/JPSJ.51.219} {\bibfield  {journal} {\bibinfo  {journal} {Journal of
  the Physical Society of Japan}\ }\textbf {\bibinfo {volume} {51}},\ \bibinfo
  {pages} {219} (\bibinfo {year} {1982})}\BibitemShut {NoStop}%
\bibitem [{\citenamefont {Maldonado}\ \emph {et~al.}(2013)\citenamefont
  {Maldonado}, \citenamefont {Vieira},\ and\ \citenamefont
  {Suderow}}]{PhysRevB.88.064518}%
  \BibitemOpen
  \bibfield  {author} {\bibinfo {author} {\bibfnamefont {A.}~\bibnamefont
  {Maldonado}}, \bibinfo {author} {\bibfnamefont {S.}~\bibnamefont {Vieira}}, \
  and\ \bibinfo {author} {\bibfnamefont {H.}~\bibnamefont {Suderow}},\
  }\bibfield  {title} {\emph {\bibinfo {title} {Supercurrent on a vortex core
  in {2H-NbSe$_2$}: Current-driven scanning tunneling spectroscopy
  measurements}},\ }\href {\doibase 10.1103/PhysRevB.88.064518} {\bibfield
  {journal} {\bibinfo  {journal} {Phys. Rev. B}\ }\textbf {\bibinfo {volume}
  {88}},\ \bibinfo {pages} {064518} (\bibinfo {year} {2013})}\BibitemShut
  {NoStop}%
\bibitem [{\citenamefont {Lindeman}\ \emph {et~al.}(2004)\citenamefont
  {Lindeman}, \citenamefont {Bandler}, \citenamefont {Brekosky}, \citenamefont
  {Chervenak}, \citenamefont {Figueroa-Feliciano}, \citenamefont {Finkbeiner},
  \citenamefont {Li},\ and\ \citenamefont {Kilbourne}}]{Lindeman2004}%
  \BibitemOpen
  \bibfield  {author} {\bibinfo {author} {\bibfnamefont {M.~A.}\ \bibnamefont
  {Lindeman}}, \bibinfo {author} {\bibfnamefont {S.}~\bibnamefont {Bandler}},
  \bibinfo {author} {\bibfnamefont {R.~P.}\ \bibnamefont {Brekosky}}, \bibinfo
  {author} {\bibfnamefont {J.~A.}\ \bibnamefont {Chervenak}}, \bibinfo {author}
  {\bibfnamefont {E.}~\bibnamefont {Figueroa-Feliciano}}, \bibinfo {author}
  {\bibfnamefont {F.~M.}\ \bibnamefont {Finkbeiner}}, \bibinfo {author}
  {\bibfnamefont {M.~J.}\ \bibnamefont {Li}}, \ and\ \bibinfo {author}
  {\bibfnamefont {C.~A.}\ \bibnamefont {Kilbourne}},\ }\bibfield  {title}
  {\emph {\bibinfo {title} {{Impedance measurements and modeling of a
  transition-edge-sensor calorimeter}}},\ }\href {\doibase 10.1063/1.1711144}
  {\bibfield  {journal} {\bibinfo  {journal} {Review of Scientific
  Instruments}\ }\textbf {\bibinfo {volume} {75}},\ \bibinfo {pages} {1283}
  (\bibinfo {year} {2004})}\BibitemShut {NoStop}%
\bibitem [{\citenamefont {Taralli}\ \emph {et~al.}(2010)\citenamefont
  {Taralli}, \citenamefont {Portesi}, \citenamefont {Lolli}, \citenamefont
  {Monticone}, \citenamefont {Rajteri}, \citenamefont {Novikov},\ and\
  \citenamefont {Beyer}}]{Taralli2010a}%
  \BibitemOpen
  \bibfield  {author} {\bibinfo {author} {\bibfnamefont {E.}~\bibnamefont
  {Taralli}}, \bibinfo {author} {\bibfnamefont {C.}~\bibnamefont {Portesi}},
  \bibinfo {author} {\bibfnamefont {L.}~\bibnamefont {Lolli}}, \bibinfo
  {author} {\bibfnamefont {E.}~\bibnamefont {Monticone}}, \bibinfo {author}
  {\bibfnamefont {M.}~\bibnamefont {Rajteri}}, \bibinfo {author} {\bibfnamefont
  {I.}~\bibnamefont {Novikov}}, \ and\ \bibinfo {author} {\bibfnamefont
  {J.}~\bibnamefont {Beyer}},\ }\bibfield  {title} {\emph {\bibinfo {title}
  {{Impedance measurements on a fast transition-edge sensor for optical and
  near-infrared range}}},\ }\href
  {http://dx.doi.org/10.1088/0953-2048/23/10/105012} {\bibfield  {journal}
  {\bibinfo  {journal} {Superconductor Science and Technology}\ }\textbf
  {\bibinfo {volume} {23}} (\bibinfo {year} {2010})}\BibitemShut {NoStop}%
\bibitem [{\citenamefont {Schmidt}(2009)}]{Schmidt2009a}%
  \BibitemOpen
  \bibfield  {author} {\bibinfo {author} {\bibfnamefont {V.~V.}\ \bibnamefont
  {Schmidt}},\ }\href@noop {} {\emph {\bibinfo {title} {Springer}}}\ (\bibinfo
  {year} {2009})\BibitemShut {NoStop}%
\bibitem [{\citenamefont {Takita}\ and\ \citenamefont
  {Masuda}(1985)}]{Takita1985}%
  \BibitemOpen
  \bibfield  {author} {\bibinfo {author} {\bibfnamefont {K.}~\bibnamefont
  {Takita}}\ and\ \bibinfo {author} {\bibfnamefont {K.}~\bibnamefont
  {Masuda}},\ }\bibfield  {title} {\emph {\bibinfo {title} {{Charge density
  wave transition and superconductivity in {2H-NbSe$_2$} . Direct measurement
  of the penetration depth in a layered superconductor}}},\ }\href {\doibase
  10.1007/BF00682569} {\bibfield  {journal} {\bibinfo  {journal} {Journal of
  Low Temperature Physics}\ }\textbf {\bibinfo {volume} {58}},\ \bibinfo
  {pages} {127} (\bibinfo {year} {1985})}\BibitemShut {NoStop}%
\bibitem [{\citenamefont {Ganta}\ \emph
  {et~al.}(2011{\natexlab{a}})\citenamefont {Ganta}, \citenamefont {Dale},
  \citenamefont {Rezac},\ and\ \citenamefont {Rosenberger}}]{ganta2011optical}%
  \BibitemOpen
  \bibfield  {author} {\bibinfo {author} {\bibfnamefont {D.}~\bibnamefont
  {Ganta}}, \bibinfo {author} {\bibfnamefont {E.}~\bibnamefont {Dale}},
  \bibinfo {author} {\bibfnamefont {J.}~\bibnamefont {Rezac}}, \ and\ \bibinfo
  {author} {\bibfnamefont {A.}~\bibnamefont {Rosenberger}},\ }\bibfield
  {title} {\emph {\bibinfo {title} {Optical method for measuring thermal
  accommodation coefficients using a whispering-gallery microresonator}},\
  }\href@noop {} {\bibfield  {journal} {\bibinfo  {journal} {The Journal of
  chemical physics}\ }\textbf {\bibinfo {volume} {135}},\ \bibinfo {pages}
  {084313} (\bibinfo {year} {2011}{\natexlab{a}})}\BibitemShut {NoStop}%
\bibitem [{\citenamefont {Huang}\ \emph {et~al.}(2007)\citenamefont {Huang},
  \citenamefont {Lin}, \citenamefont {Chang}, \citenamefont {Sun},
  \citenamefont {Shen}, \citenamefont {Chou}, \citenamefont {Berger},
  \citenamefont {Lee},\ and\ \citenamefont {Yang}}]{Huang2007a}%
  \BibitemOpen
  \bibfield  {author} {\bibinfo {author} {\bibfnamefont {C.~L.}\ \bibnamefont
  {Huang}}, \bibinfo {author} {\bibfnamefont {J.-Y.}\ \bibnamefont {Lin}},
  \bibinfo {author} {\bibfnamefont {Y.~T.}\ \bibnamefont {Chang}}, \bibinfo
  {author} {\bibfnamefont {C.~P.}\ \bibnamefont {Sun}}, \bibinfo {author}
  {\bibfnamefont {H.~Y.}\ \bibnamefont {Shen}}, \bibinfo {author}
  {\bibfnamefont {C.~C.}\ \bibnamefont {Chou}}, \bibinfo {author}
  {\bibfnamefont {H.}~\bibnamefont {Berger}}, \bibinfo {author} {\bibfnamefont
  {T.~K.}\ \bibnamefont {Lee}}, \ and\ \bibinfo {author} {\bibfnamefont
  {H.~D.}\ \bibnamefont {Yang}},\ }\bibfield  {title} {\emph {\bibinfo {title}
  {{Experimental evidence for a two-gap structure of superconducting {NbSe$_2$}
  specific-heat study in external magnetic fields}}},\ }\href {\doibase
  10.1103/PhysRevB.76.212504} {\bibfield  {journal} {\bibinfo  {journal} {Phys.
  Rev. B}\ }\textbf {\bibinfo {volume} {76}},\ \bibinfo {pages} {212504}
  (\bibinfo {year} {2007})}\BibitemShut {NoStop}%
\bibitem [{\citenamefont {Boaknin}\ \emph {et~al.}(2003)\citenamefont
  {Boaknin}, \citenamefont {Tanatar}, \citenamefont {Paglione}, \citenamefont
  {Hawthorn}, \citenamefont {Ronning}, \citenamefont {Hill}, \citenamefont
  {Sutherland}, \citenamefont {Taillefer}, \citenamefont {Sonier},
  \citenamefont {Hayden},\ and\ \citenamefont {Brill}}]{PhysRevLett.90.117003}%
  \BibitemOpen
  \bibfield  {author} {\bibinfo {author} {\bibfnamefont {E.}~\bibnamefont
  {Boaknin}}, \bibinfo {author} {\bibfnamefont {M.~A.}\ \bibnamefont
  {Tanatar}}, \bibinfo {author} {\bibfnamefont {J.}~\bibnamefont {Paglione}},
  \bibinfo {author} {\bibfnamefont {D.}~\bibnamefont {Hawthorn}}, \bibinfo
  {author} {\bibfnamefont {F.}~\bibnamefont {Ronning}}, \bibinfo {author}
  {\bibfnamefont {R.~W.}\ \bibnamefont {Hill}}, \bibinfo {author}
  {\bibfnamefont {M.}~\bibnamefont {Sutherland}}, \bibinfo {author}
  {\bibfnamefont {L.}~\bibnamefont {Taillefer}}, \bibinfo {author}
  {\bibfnamefont {J.}~\bibnamefont {Sonier}}, \bibinfo {author} {\bibfnamefont
  {S.~M.}\ \bibnamefont {Hayden}}, \ and\ \bibinfo {author} {\bibfnamefont
  {J.~W.}\ \bibnamefont {Brill}},\ }\bibfield  {title} {\emph {\bibinfo {title}
  {Heat conduction in the vortex state of $nbse_2$: Evidence for multiband
  superconductivity}},\ }\href {\doibase 10.1103/PhysRevLett.90.117003}
  {\bibfield  {journal} {\bibinfo  {journal} {Phys. Rev. Lett.}\ }\textbf
  {\bibinfo {volume} {90}},\ \bibinfo {pages} {117003} (\bibinfo {year}
  {2003})}\BibitemShut {NoStop}%
\bibitem [{\citenamefont {Blatter}\ \emph {et~al.}(1994)\citenamefont
  {Blatter}, \citenamefont {Feigel'man}, \citenamefont {Geshkenbein},
  \citenamefont {Larkin},\ and\ \citenamefont {Vinokur}}]{Blatter1994}%
  \BibitemOpen
  \bibfield  {author} {\bibinfo {author} {\bibfnamefont {G.}~\bibnamefont
  {Blatter}}, \bibinfo {author} {\bibfnamefont {M.~V.}\ \bibnamefont
  {Feigel'man}}, \bibinfo {author} {\bibfnamefont {V.~B.}\ \bibnamefont
  {Geshkenbein}}, \bibinfo {author} {\bibfnamefont {A.~I.}\ \bibnamefont
  {Larkin}}, \ and\ \bibinfo {author} {\bibfnamefont {V.~M.}\ \bibnamefont
  {Vinokur}},\ }\bibfield  {title} {\emph {\bibinfo {title} {Vortices in
  high-temperature superconductors}},\ }\href {\doibase
  10.1103/RevModPhys.66.1125} {\bibfield  {journal} {\bibinfo  {journal} {Rev.
  Mod. Phys.}\ }\textbf {\bibinfo {volume} {66}},\ \bibinfo {pages} {1125}
  (\bibinfo {year} {1994})}\BibitemShut {NoStop}%
\bibitem [{\citenamefont {Kim}\ \emph {et~al.}(1965)\citenamefont {Kim},
  \citenamefont {Hempstead},\ and\ \citenamefont {Strnad}}]{PhysRev.139.A1163}%
  \BibitemOpen
  \bibfield  {author} {\bibinfo {author} {\bibfnamefont {Y.~B.}\ \bibnamefont
  {Kim}}, \bibinfo {author} {\bibfnamefont {C.~F.}\ \bibnamefont {Hempstead}},
  \ and\ \bibinfo {author} {\bibfnamefont {A.~R.}\ \bibnamefont {Strnad}},\
  }\bibfield  {title} {\emph {\bibinfo {title} {Flux-flow resistance in
  type-{II} superconductors}},\ }\href {\doibase 10.1103/PhysRev.139.A1163}
  {\bibfield  {journal} {\bibinfo  {journal} {Phys. Rev.}\ }\textbf {\bibinfo
  {volume} {139}},\ \bibinfo {pages} {A1163} (\bibinfo {year}
  {1965})}\BibitemShut {NoStop}%
\bibitem [{\citenamefont {Maki}(1971)}]{MAKI1971124}%
  \BibitemOpen
  \bibfield  {author} {\bibinfo {author} {\bibfnamefont {K.}~\bibnamefont
  {Maki}},\ }\bibfield  {title} {\emph {\bibinfo {title} {Vortex motion in
  superconductors}},\ }\href {\doibase
  https://doi.org/10.1016/0031-8914(71)90247-3} {\bibfield  {journal} {\bibinfo
   {journal} {Physica}\ }\textbf {\bibinfo {volume} {55}},\ \bibinfo {pages}
  {124 } (\bibinfo {year} {1971})}\BibitemShut {NoStop}%
\bibitem [{\citenamefont {Bardeen}\ and\ \citenamefont
  {Stephen}(1965)}]{PhysRev.140.A1197}%
  \BibitemOpen
  \bibfield  {author} {\bibinfo {author} {\bibfnamefont {J.}~\bibnamefont
  {Bardeen}}\ and\ \bibinfo {author} {\bibfnamefont {M.~J.}\ \bibnamefont
  {Stephen}},\ }\bibfield  {title} {\emph {\bibinfo {title} {Theory of the
  motion of vortices in superconductors}},\ }\href {\doibase
  10.1103/PhysRev.140.A1197} {\bibfield  {journal} {\bibinfo  {journal} {Phys.
  Rev.}\ }\textbf {\bibinfo {volume} {140}},\ \bibinfo {pages} {A1197}
  (\bibinfo {year} {1965})}\BibitemShut {NoStop}%
\bibitem [{\citenamefont {Hu}\ and\ \citenamefont
  {Thompson}(1972)}]{PhysRevB.6.110}%
  \BibitemOpen
  \bibfield  {author} {\bibinfo {author} {\bibfnamefont {C.-R.}\ \bibnamefont
  {Hu}}\ and\ \bibinfo {author} {\bibfnamefont {R.~S.}\ \bibnamefont
  {Thompson}},\ }\bibfield  {title} {\emph {\bibinfo {title} {Dynamic structure
  of vortices in superconductors. {II}. ${H}\ensuremath{\ll}{H}_{c2}$}},\
  }\href {\doibase 10.1103/PhysRevB.6.110} {\bibfield  {journal} {\bibinfo
  {journal} {Phys. Rev. B}\ }\textbf {\bibinfo {volume} {6}},\ \bibinfo {pages}
  {110} (\bibinfo {year} {1972})}\BibitemShut {NoStop}%
\bibitem [{\citenamefont {Clem}(1968)}]{PhysRevLett.20.735}%
  \BibitemOpen
  \bibfield  {author} {\bibinfo {author} {\bibfnamefont {J.~R.}\ \bibnamefont
  {Clem}},\ }\bibfield  {title} {\emph {\bibinfo {title} {Local
  temperature-gradient contribution to flux-flow viscosity in
  superconductors}},\ }\href {\doibase 10.1103/PhysRevLett.20.735} {\bibfield
  {journal} {\bibinfo  {journal} {Phys. Rev. Lett.}\ }\textbf {\bibinfo
  {volume} {20}},\ \bibinfo {pages} {735} (\bibinfo {year} {1968})}\BibitemShut
  {NoStop}%
\bibitem [{\citenamefont {Kogan}(2018)}]{PhysRevB.97.094510}%
  \BibitemOpen
  \bibfield  {author} {\bibinfo {author} {\bibfnamefont {V.~G.}\ \bibnamefont
  {Kogan}},\ }\bibfield  {title} {\emph {\bibinfo {title} {Time-dependent
  london approach: Dissipation due to out-of-core normal excitations by moving
  vortices}},\ }\href {\doibase 10.1103/PhysRevB.97.094510} {\bibfield
  {journal} {\bibinfo  {journal} {Phys. Rev. B}\ }\textbf {\bibinfo {volume}
  {97}},\ \bibinfo {pages} {094510} (\bibinfo {year} {2018})}\BibitemShut
  {NoStop}%
\bibitem [{\citenamefont {Embon}\ \emph {et~al.}(2017)\citenamefont {Embon},
  \citenamefont {Anahory}, \citenamefont {Jelic}, \citenamefont {Lachman},
  \citenamefont {Myasoedov}, \citenamefont {Huber}, \citenamefont {Mikitik},
  \citenamefont {Silhanek}, \citenamefont {Milosevic}, \citenamefont
  {Gurevich},\ and\ \citenamefont {Zeldov}}]{Embon2017}%
  \BibitemOpen
  \bibfield  {author} {\bibinfo {author} {\bibfnamefont {L.}~\bibnamefont
  {Embon}}, \bibinfo {author} {\bibfnamefont {Y.}~\bibnamefont {Anahory}},
  \bibinfo {author} {\bibfnamefont {Z.~L.}\ \bibnamefont {Jelic}}, \bibinfo
  {author} {\bibfnamefont {E.~O.}\ \bibnamefont {Lachman}}, \bibinfo {author}
  {\bibfnamefont {Y.}~\bibnamefont {Myasoedov}}, \bibinfo {author}
  {\bibfnamefont {M.~E.}\ \bibnamefont {Huber}}, \bibinfo {author}
  {\bibfnamefont {G.~P.}\ \bibnamefont {Mikitik}}, \bibinfo {author}
  {\bibfnamefont {A.~V.}\ \bibnamefont {Silhanek}}, \bibinfo {author}
  {\bibfnamefont {M.~V.}\ \bibnamefont {Milosevic}}, \bibinfo {author}
  {\bibfnamefont {A.}~\bibnamefont {Gurevich}}, \ and\ \bibinfo {author}
  {\bibfnamefont {E.}~\bibnamefont {Zeldov}},\ }\bibfield  {title} {\emph
  {\bibinfo {title} {Imaging of super-fast dynamics and flow instabilities of
  superconducting vortices}},\ }\href {\doibase 10.1038/s41467-017-00089-3}
  {\bibfield  {journal} {\bibinfo  {journal} {Nature Communications}\ }\textbf
  {\bibinfo {volume} {8}},\ \bibinfo {pages} {85} (\bibinfo {year}
  {2017})}\BibitemShut {NoStop}%
\bibitem [{\citenamefont {Grimaldi}\ \emph {et~al.}(2015)\citenamefont
  {Grimaldi}, \citenamefont {Leo}, \citenamefont {Sabatino}, \citenamefont
  {Carapella}, \citenamefont {Nigro}, \citenamefont {Pace}, \citenamefont
  {Moshchalkov},\ and\ \citenamefont {Silhanek}}]{PhysRevB.92.024513}%
  \BibitemOpen
  \bibfield  {author} {\bibinfo {author} {\bibfnamefont {G.}~\bibnamefont
  {Grimaldi}}, \bibinfo {author} {\bibfnamefont {A.}~\bibnamefont {Leo}},
  \bibinfo {author} {\bibfnamefont {P.}~\bibnamefont {Sabatino}}, \bibinfo
  {author} {\bibfnamefont {G.}~\bibnamefont {Carapella}}, \bibinfo {author}
  {\bibfnamefont {A.}~\bibnamefont {Nigro}}, \bibinfo {author} {\bibfnamefont
  {S.}~\bibnamefont {Pace}}, \bibinfo {author} {\bibfnamefont {V.~V.}\
  \bibnamefont {Moshchalkov}}, \ and\ \bibinfo {author} {\bibfnamefont {A.~V.}\
  \bibnamefont {Silhanek}},\ }\bibfield  {title} {\emph {\bibinfo {title}
  {Speed limit to the abrikosov lattice in mesoscopic superconductors}},\
  }\href {\doibase 10.1103/PhysRevB.92.024513} {\bibfield  {journal} {\bibinfo
  {journal} {Phys. Rev. B}\ }\textbf {\bibinfo {volume} {92}},\ \bibinfo
  {pages} {024513} (\bibinfo {year} {2015})}\BibitemShut {NoStop}%
\bibitem [{\citenamefont {Dobrovolskiy}\ \emph {et~al.}(2019)\citenamefont
  {Dobrovolskiy}, \citenamefont {Bevz}, \citenamefont {Begun}, \citenamefont
  {Sachser}, \citenamefont {Vovk},\ and\ \citenamefont
  {Huth}}]{PhysRevApplied.11.054064}%
  \BibitemOpen
  \bibfield  {author} {\bibinfo {author} {\bibfnamefont {O.}~\bibnamefont
  {Dobrovolskiy}}, \bibinfo {author} {\bibfnamefont {V.}~\bibnamefont {Bevz}},
  \bibinfo {author} {\bibfnamefont {E.}~\bibnamefont {Begun}}, \bibinfo
  {author} {\bibfnamefont {R.}~\bibnamefont {Sachser}}, \bibinfo {author}
  {\bibfnamefont {R.}~\bibnamefont {Vovk}}, \ and\ \bibinfo {author}
  {\bibfnamefont {M.}~\bibnamefont {Huth}},\ }\bibfield  {title} {\emph
  {\bibinfo {title} {Fast dynamics of guided magnetic flux quanta}},\ }\href
  {\doibase 10.1103/PhysRevApplied.11.054064} {\bibfield  {journal} {\bibinfo
  {journal} {Phys. Rev. Applied}\ }\textbf {\bibinfo {volume} {11}},\ \bibinfo
  {pages} {054064} (\bibinfo {year} {2019})}\BibitemShut {NoStop}%
\bibitem [{\citenamefont {Estell{\'e}s-Duart}\ \emph
  {et~al.}(2018)\citenamefont {Estell{\'e}s-Duart}, \citenamefont {Ortu{\~n}o},
  \citenamefont {Somoza}, \citenamefont {Vinokur},\ and\ \citenamefont
  {Gurevich}}]{Estelles-Duart2018}%
  \BibitemOpen
  \bibfield  {author} {\bibinfo {author} {\bibfnamefont {F.}~\bibnamefont
  {Estell{\'e}s-Duart}}, \bibinfo {author} {\bibfnamefont {M.}~\bibnamefont
  {Ortu{\~n}o}}, \bibinfo {author} {\bibfnamefont {A.~M.}\ \bibnamefont
  {Somoza}}, \bibinfo {author} {\bibfnamefont {V.~M.}\ \bibnamefont {Vinokur}},
  \ and\ \bibinfo {author} {\bibfnamefont {A.}~\bibnamefont {Gurevich}},\
  }\bibfield  {title} {\emph {\bibinfo {title} {Current-driven production of
  vortex-antivortex pairs in planar josephson junction arrays and phase cracks
  in long-range order}},\ }\href {\doibase 10.1038/s41598-018-33467-y}
  {\bibfield  {journal} {\bibinfo  {journal} {Scientific Reports}\ }\textbf
  {\bibinfo {volume} {8}},\ \bibinfo {pages} {15460} (\bibinfo {year}
  {2018})}\BibitemShut {NoStop}%
\bibitem [{\citenamefont {Jing}\ \emph {et~al.}(2018)\citenamefont {Jing},
  \citenamefont {Yong},\ and\ \citenamefont {Zhou}}]{Jing_2018}%
  \BibitemOpen
  \bibfield  {author} {\bibinfo {author} {\bibfnamefont {Z.}~\bibnamefont
  {Jing}}, \bibinfo {author} {\bibfnamefont {H.}~\bibnamefont {Yong}}, \ and\
  \bibinfo {author} {\bibfnamefont {Y.}~\bibnamefont {Zhou}},\ }\bibfield
  {title} {\emph {\bibinfo {title} {Thermal coupling effect on the vortex
  dynamics of superconducting thin films: time-dependent
  ginzburg{\textendash}landau simulations}},\ }\href {\doibase
  10.1088/1361-6668/aab3be} {\bibfield  {journal} {\bibinfo  {journal}
  {Superconductor Science and Technology}\ }\textbf {\bibinfo {volume} {31}},\
  \bibinfo {pages} {055007} (\bibinfo {year} {2018})}\BibitemShut {NoStop}%
\bibitem [{\citenamefont {Leroux}\ \emph {et~al.}(2019)\citenamefont {Leroux},
  \citenamefont {Balakirev}, \citenamefont {Miura}, \citenamefont {Agatsuma},
  \citenamefont {Civale},\ and\ \citenamefont
  {Maiorov}}]{PhysRevApplied.11.054005}%
  \BibitemOpen
  \bibfield  {author} {\bibinfo {author} {\bibfnamefont {M.}~\bibnamefont
  {Leroux}}, \bibinfo {author} {\bibfnamefont {F.~F.}\ \bibnamefont
  {Balakirev}}, \bibinfo {author} {\bibfnamefont {M.}~\bibnamefont {Miura}},
  \bibinfo {author} {\bibfnamefont {K.}~\bibnamefont {Agatsuma}}, \bibinfo
  {author} {\bibfnamefont {L.}~\bibnamefont {Civale}}, \ and\ \bibinfo {author}
  {\bibfnamefont {B.}~\bibnamefont {Maiorov}},\ }\bibfield  {title} {\emph
  {\bibinfo {title} {Dynamics and critical currents in fast superconducting
  vortices at high pulsed magnetic fields}},\ }\href {\doibase
  10.1103/PhysRevApplied.11.054005} {\bibfield  {journal} {\bibinfo  {journal}
  {Phys. Rev. Applied}\ }\textbf {\bibinfo {volume} {11}},\ \bibinfo {pages}
  {054005} (\bibinfo {year} {2019})}\BibitemShut {NoStop}%
\bibitem [{\citenamefont {Kalisky}\ \emph {et~al.}(2006)\citenamefont
  {Kalisky}, \citenamefont {Aronov}, \citenamefont {Koren}, \citenamefont
  {Shaulov}, \citenamefont {Yeshurun},\ and\ \citenamefont
  {Huebener}}]{PhysRevLett.97.067003}%
  \BibitemOpen
  \bibfield  {author} {\bibinfo {author} {\bibfnamefont {B.}~\bibnamefont
  {Kalisky}}, \bibinfo {author} {\bibfnamefont {P.}~\bibnamefont {Aronov}},
  \bibinfo {author} {\bibfnamefont {G.}~\bibnamefont {Koren}}, \bibinfo
  {author} {\bibfnamefont {A.}~\bibnamefont {Shaulov}}, \bibinfo {author}
  {\bibfnamefont {Y.}~\bibnamefont {Yeshurun}}, \ and\ \bibinfo {author}
  {\bibfnamefont {R.~P.}\ \bibnamefont {Huebener}},\ }\bibfield  {title} {\emph
  {\bibinfo {title} {Flux-flow resistivity anisotropy in the instability regime
  of the $a\mathrm{\text{\ensuremath{-}}}b$ plane of epitaxial superconducting
  {YBa$_2$Cu$_3$0$_7$} thin films}},\ }\href {\doibase
  10.1103/PhysRevLett.97.067003} {\bibfield  {journal} {\bibinfo  {journal}
  {Phys. Rev. Lett.}\ }\textbf {\bibinfo {volume} {97}},\ \bibinfo {pages}
  {067003} (\bibinfo {year} {2006})}\BibitemShut {NoStop}%
\bibitem [{\citenamefont {Kunchur}\ \emph {et~al.}(1994)\citenamefont
  {Kunchur}, \citenamefont {Christen}, \citenamefont {Klabunde},\ and\
  \citenamefont {Phillips}}]{PhysRevLett.72.752}%
  \BibitemOpen
  \bibfield  {author} {\bibinfo {author} {\bibfnamefont {M.~N.}\ \bibnamefont
  {Kunchur}}, \bibinfo {author} {\bibfnamefont {D.~K.}\ \bibnamefont
  {Christen}}, \bibinfo {author} {\bibfnamefont {C.~E.}\ \bibnamefont
  {Klabunde}}, \ and\ \bibinfo {author} {\bibfnamefont {J.~M.}\ \bibnamefont
  {Phillips}},\ }\bibfield  {title} {\emph {\bibinfo {title} {Pair-breaking
  effect of high current densities on the superconducting transition in
  {YBa$_2$Cu$_3$O$_7$}}},\ }\href {\doibase 10.1103/PhysRevLett.72.752}
  {\bibfield  {journal} {\bibinfo  {journal} {Phys. Rev. Lett.}\ }\textbf
  {\bibinfo {volume} {72}},\ \bibinfo {pages} {752} (\bibinfo {year}
  {1994})}\BibitemShut {NoStop}%
\bibitem [{\citenamefont {Benyamini}\ \emph {et~al.}(2019)\citenamefont
  {Benyamini}, \citenamefont {Kennes}, \citenamefont {Telford}, \citenamefont
  {Watanabe}, \citenamefont {Taniguchi}, \citenamefont {Millis}, \citenamefont
  {Hone}, \citenamefont {Dean},\ and\ \citenamefont
  {Pasupathy}}]{benyamini2019blockade}%
  \BibitemOpen
  \bibfield  {author} {\bibinfo {author} {\bibfnamefont {A.}~\bibnamefont
  {Benyamini}}, \bibinfo {author} {\bibfnamefont {D.~M.}\ \bibnamefont
  {Kennes}}, \bibinfo {author} {\bibfnamefont {E.}~\bibnamefont {Telford}},
  \bibinfo {author} {\bibfnamefont {K.}~\bibnamefont {Watanabe}}, \bibinfo
  {author} {\bibfnamefont {T.}~\bibnamefont {Taniguchi}}, \bibinfo {author}
  {\bibfnamefont {A.}~\bibnamefont {Millis}}, \bibinfo {author} {\bibfnamefont
  {J.}~\bibnamefont {Hone}}, \bibinfo {author} {\bibfnamefont {C.~R.}\
  \bibnamefont {Dean}}, \ and\ \bibinfo {author} {\bibfnamefont
  {A.}~\bibnamefont {Pasupathy}},\ }\href@noop {} {\emph {\bibinfo {title}
  {Blockade of vortex flow by thermal fluctuations in atomically thin
  clean-limit superconductors}}} (\bibinfo {year} {2019}),\ \Eprint
  {http://arxiv.org/abs/1909.08469} {arXiv:1909.08469 [cond-mat.mes-hall]}
  \BibitemShut {NoStop}%
\bibitem [{\citenamefont {Sata}\ \emph {et~al.}(2018)\citenamefont {Sata},
  \citenamefont {Moriya}, \citenamefont {Yabuki}, \citenamefont {Masubuchi},\
  and\ \citenamefont {Machida}}]{PhysRevB.98.035422}%
  \BibitemOpen
  \bibfield  {author} {\bibinfo {author} {\bibfnamefont {Y.}~\bibnamefont
  {Sata}}, \bibinfo {author} {\bibfnamefont {R.}~\bibnamefont {Moriya}},
  \bibinfo {author} {\bibfnamefont {N.}~\bibnamefont {Yabuki}}, \bibinfo
  {author} {\bibfnamefont {S.}~\bibnamefont {Masubuchi}}, \ and\ \bibinfo
  {author} {\bibfnamefont {T.}~\bibnamefont {Machida}},\ }\bibfield  {title}
  {\emph {\bibinfo {title} {Heat transfer at the van der waals interface
  between graphene and {NbSe$_2$}}},\ }\href {\doibase
  10.1103/PhysRevB.98.035422} {\bibfield  {journal} {\bibinfo  {journal} {Phys.
  Rev. B}\ }\textbf {\bibinfo {volume} {98}},\ \bibinfo {pages} {035422}
  (\bibinfo {year} {2018})}\BibitemShut {NoStop}%
\bibitem [{\citenamefont {Funke}\ and\ \citenamefont
  {Haberstroh}(2015)}]{Funke2015a}%
  \BibitemOpen
  \bibfield  {author} {\bibinfo {author} {\bibfnamefont {T.}~\bibnamefont
  {Funke}}\ and\ \bibinfo {author} {\bibfnamefont {C.}~\bibnamefont
  {Haberstroh}},\ }\bibfield  {title} {\emph {\bibinfo {title} {{New
  measurements of multilayer insulation at variable cold temperature and
  elevated residual gas pressure}}},\ }\href
  {https://doi.org/10.1088/1757-899X/101/1/012058} {\bibfield  {journal}
  {\bibinfo  {journal} {IOP Conference Series: Materials Science and
  Engineering}\ }\textbf {\bibinfo {volume} {101}} (\bibinfo {year}
  {2015})}\BibitemShut {NoStop}%
\bibitem [{\citenamefont {Ganta}\ \emph
  {et~al.}(2011{\natexlab{b}})\citenamefont {Ganta}, \citenamefont {Dale},
  \citenamefont {Rezac},\ and\ \citenamefont {Rosenberger}}]{Ganta2011a}%
  \BibitemOpen
  \bibfield  {author} {\bibinfo {author} {\bibfnamefont {D.}~\bibnamefont
  {Ganta}}, \bibinfo {author} {\bibfnamefont {E.~B.}\ \bibnamefont {Dale}},
  \bibinfo {author} {\bibfnamefont {J.~P.}\ \bibnamefont {Rezac}}, \ and\
  \bibinfo {author} {\bibfnamefont {A.~T.}\ \bibnamefont {Rosenberger}},\
  }\bibfield  {title} {\emph {\bibinfo {title} {{Measuring thermal
  accommodation coefficients using a whispering-gallery optical
  microresonator}}},\ }\href {https://doi.org/10.1063/1.3631342} {\bibfield
  {journal} {\bibinfo  {journal} {The Journal of Chemical Physics}\ }\textbf
  {\bibinfo {volume} {135}},\ \bibinfo {pages} {435} (\bibinfo {year}
  {2011}{\natexlab{b}})}\BibitemShut {NoStop}%
\bibitem [{\citenamefont {Halbertal}\ \emph {et~al.}(2016)\citenamefont
  {Halbertal}, \citenamefont {Cuppens}, \citenamefont {Shalom}, \citenamefont
  {Embon}, \citenamefont {Shadmi}, \citenamefont {Anahory}, \citenamefont
  {Naren}, \citenamefont {Sarkar}, \citenamefont {Uri}, \citenamefont {Ronen},
  \citenamefont {Myasoedov}, \citenamefont {Levitov}, \citenamefont
  {Joselevich}, \citenamefont {Geim},\ and\ \citenamefont
  {Zeldov}}]{Halbertal2016}%
  \BibitemOpen
  \bibfield  {author} {\bibinfo {author} {\bibfnamefont {D.}~\bibnamefont
  {Halbertal}}, \bibinfo {author} {\bibfnamefont {J.}~\bibnamefont {Cuppens}},
  \bibinfo {author} {\bibfnamefont {M.~B.}\ \bibnamefont {Shalom}}, \bibinfo
  {author} {\bibfnamefont {L.}~\bibnamefont {Embon}}, \bibinfo {author}
  {\bibfnamefont {N.}~\bibnamefont {Shadmi}}, \bibinfo {author} {\bibfnamefont
  {Y.}~\bibnamefont {Anahory}}, \bibinfo {author} {\bibfnamefont {H.~R.}\
  \bibnamefont {Naren}}, \bibinfo {author} {\bibfnamefont {J.}~\bibnamefont
  {Sarkar}}, \bibinfo {author} {\bibfnamefont {A.}~\bibnamefont {Uri}},
  \bibinfo {author} {\bibfnamefont {Y.}~\bibnamefont {Ronen}}, \bibinfo
  {author} {\bibfnamefont {Y.}~\bibnamefont {Myasoedov}}, \bibinfo {author}
  {\bibfnamefont {L.~S.}\ \bibnamefont {Levitov}}, \bibinfo {author}
  {\bibfnamefont {E.}~\bibnamefont {Joselevich}}, \bibinfo {author}
  {\bibfnamefont {A.~K.}\ \bibnamefont {Geim}}, \ and\ \bibinfo {author}
  {\bibfnamefont {E.}~\bibnamefont {Zeldov}},\ }\bibfield  {title} {\emph
  {\bibinfo {title} {{Nanoscale thermal imaging of dissipation in quantum
  systems}}},\ }\href {\doibase 10.1038/nature19843} {\bibfield  {journal}
  {\bibinfo  {journal} {Nature}\ }\textbf {\bibinfo {volume} {539}},\ \bibinfo
  {pages} {407} (\bibinfo {year} {2016})}\BibitemShut {NoStop}%
\bibitem [{\citenamefont {Halbertal}\ \emph {et~al.}(2017)\citenamefont
  {Halbertal}, \citenamefont {Uri}, \citenamefont {Bagani}, \citenamefont
  {Meltzer}, \citenamefont {Marcus}, \citenamefont {Myasoedov}, \citenamefont
  {Zeldov}, \citenamefont {Shalom}, \citenamefont {Birkbeck}, \citenamefont
  {Geim},\ and\ \citenamefont {Levitov}}]{Halbertal2017}%
  \BibitemOpen
  \bibfield  {author} {\bibinfo {author} {\bibfnamefont {D.}~\bibnamefont
  {Halbertal}}, \bibinfo {author} {\bibfnamefont {A.}~\bibnamefont {Uri}},
  \bibinfo {author} {\bibfnamefont {K.}~\bibnamefont {Bagani}}, \bibinfo
  {author} {\bibfnamefont {A.~Y.}\ \bibnamefont {Meltzer}}, \bibinfo {author}
  {\bibfnamefont {I.}~\bibnamefont {Marcus}}, \bibinfo {author} {\bibfnamefont
  {Y.}~\bibnamefont {Myasoedov}}, \bibinfo {author} {\bibfnamefont
  {E.}~\bibnamefont {Zeldov}}, \bibinfo {author} {\bibfnamefont {M.~B.}\
  \bibnamefont {Shalom}}, \bibinfo {author} {\bibfnamefont {J.}~\bibnamefont
  {Birkbeck}}, \bibinfo {author} {\bibfnamefont {A.~K.}\ \bibnamefont {Geim}},
  \ and\ \bibinfo {author} {\bibfnamefont {L.~S.}\ \bibnamefont {Levitov}},\
  }\bibfield  {title} {\emph {\bibinfo {title} {{Imaging resonant dissipation
  from individual atomic defects in graphene}}},\ }\href {\doibase
  10.1126/science.aan0877} {\bibfield  {journal} {\bibinfo  {journal}
  {Science}\ }\textbf {\bibinfo {volume} {358}},\ \bibinfo {pages} {1303}
  (\bibinfo {year} {2017})}\BibitemShut {NoStop}%
\bibitem [{\citenamefont {Miller}\ and\ \citenamefont
  {Pierce}(1973)}]{PhysRevB.8.4164}%
  \BibitemOpen
  \bibfield  {author} {\bibinfo {author} {\bibfnamefont {J.~R.}\ \bibnamefont
  {Miller}}\ and\ \bibinfo {author} {\bibfnamefont {J.~M.}\ \bibnamefont
  {Pierce}},\ }\bibfield  {title} {\emph {\bibinfo {title} {Fluctuation effects
  in the complex impedance of superconducting tin-whisker crystals near
  ${T}_{c}$}},\ }\href {\doibase 10.1103/PhysRevB.8.4164} {\bibfield  {journal}
  {\bibinfo  {journal} {Phys. Rev. B}\ }\textbf {\bibinfo {volume} {8}},\
  \bibinfo {pages} {4164} (\bibinfo {year} {1973})}\BibitemShut {NoStop}%
\bibitem [{\citenamefont {Kremen}\ \emph {et~al.}(2018)\citenamefont {Kremen},
  \citenamefont {Khan}, \citenamefont {Loh}, \citenamefont {Baturina},
  \citenamefont {Trivedi}, \citenamefont {Frydman},\ and\ \citenamefont
  {Kalisky}}]{Kremen2018}%
  \BibitemOpen
  \bibfield  {author} {\bibinfo {author} {\bibfnamefont {A.}~\bibnamefont
  {Kremen}}, \bibinfo {author} {\bibfnamefont {H.}~\bibnamefont {Khan}},
  \bibinfo {author} {\bibfnamefont {Y.~L.}\ \bibnamefont {Loh}}, \bibinfo
  {author} {\bibfnamefont {T.~I.}\ \bibnamefont {Baturina}}, \bibinfo {author}
  {\bibfnamefont {N.}~\bibnamefont {Trivedi}}, \bibinfo {author} {\bibfnamefont
  {A.}~\bibnamefont {Frydman}}, \ and\ \bibinfo {author} {\bibfnamefont
  {B.}~\bibnamefont {Kalisky}},\ }\bibfield  {title} {\emph {\bibinfo {title}
  {Imaging quantum fluctuations near criticality}},\ }\href {\doibase
  10.1038/s41567-018-0264-z} {\bibfield  {journal} {\bibinfo  {journal} {Nature
  Physics}\ }\textbf {\bibinfo {volume} {14}},\ \bibinfo {pages} {1205}
  (\bibinfo {year} {2018})}\BibitemShut {NoStop}%
\bibitem [{\citenamefont {Larkin}\ and\ \citenamefont
  {Varlamov}(2008)}]{Larkin2008}%
  \BibitemOpen
  \bibfield  {author} {\bibinfo {author} {\bibfnamefont {A.~I.}\ \bibnamefont
  {Larkin}}\ and\ \bibinfo {author} {\bibfnamefont {A.~A.}\ \bibnamefont
  {Varlamov}},\ }\emph {\bibinfo {title} {Fluctuation phenomena in
  superconductors}},\ in\ \href {\doibase 10.1007/978-3-540-73253-2_10} {\emph
  {\bibinfo {booktitle} {Superconductivity: Conventional and Unconventional
  Superconductors}}},\ \bibinfo {editor} {edited by\ \bibinfo {editor}
  {\bibfnamefont {K.~H.}\ \bibnamefont {Bennemann}}\ and\ \bibinfo {editor}
  {\bibfnamefont {J.~B.}\ \bibnamefont {Ketterson}}}\ (\bibinfo  {publisher}
  {Springer Berlin Heidelberg},\ \bibinfo {address} {Berlin, Heidelberg},\
  \bibinfo {year} {2008})\ pp.\ \bibinfo {pages} {369--458}\BibitemShut
  {NoStop}%
\bibitem [{\citenamefont {Larkin}\ and\ \citenamefont
  {Ovchinnikov}(2001)}]{Larkin2001}%
  \BibitemOpen
  \bibfield  {author} {\bibinfo {author} {\bibfnamefont {A.~I.}\ \bibnamefont
  {Larkin}}\ and\ \bibinfo {author} {\bibfnamefont {Y.~N.}\ \bibnamefont
  {Ovchinnikov}},\ }\bibfield  {title} {\emph {\bibinfo {title} {Nonlinear
  fluctuation phenomena in the transport properties of superconductors}},\
  }\href {\doibase 10.1134/1.1364749} {\bibfield  {journal} {\bibinfo
  {journal} {Journal of Experimental and Theoretical Physics}\ }\textbf
  {\bibinfo {volume} {92}},\ \bibinfo {pages} {519} (\bibinfo {year}
  {2001})}\BibitemShut {NoStop}%
\bibitem [{\citenamefont {Maki}(1968)}]{10.1143/PTP.39.897}%
  \BibitemOpen
  \bibfield  {author} {\bibinfo {author} {\bibfnamefont {K.}~\bibnamefont
  {Maki}},\ }\bibfield  {title} {\emph {\bibinfo {title} {{The Critical
  Fluctuation of the Order Parameter in Type-II Superconductors}}},\ }\href
  {https://doi.org/10.1143/PTP.39.897} {\bibfield  {journal} {\bibinfo
  {journal} {Progress of Theoretical Physics}\ }\textbf {\bibinfo {volume}
  {39}},\ \bibinfo {pages} {897} (\bibinfo {year} {1968})}\BibitemShut
  {NoStop}%
\bibitem [{\citenamefont {Thompson}(1970)}]{PhysRevB.1.327}%
  \BibitemOpen
  \bibfield  {author} {\bibinfo {author} {\bibfnamefont {R.~S.}\ \bibnamefont
  {Thompson}},\ }\bibfield  {title} {\emph {\bibinfo {title} {Microwave, flux
  flow, and fluctuation resistance of dirty type-ii superconductors}},\ }\href
  {\doibase 10.1103/PhysRevB.1.327} {\bibfield  {journal} {\bibinfo  {journal}
  {Phys. Rev. B}\ }\textbf {\bibinfo {volume} {1}},\ \bibinfo {pages} {327}
  (\bibinfo {year} {1970})}\BibitemShut {NoStop}%
\bibitem [{\citenamefont {Carballeira}\ \emph {et~al.}(2000)\citenamefont
  {Carballeira}, \citenamefont {Mosqueira}, \citenamefont {Revcolevschi},\ and\
  \citenamefont {Vidal}}]{Carballeira2000}%
  \BibitemOpen
  \bibfield  {author} {\bibinfo {author} {\bibfnamefont {C.}~\bibnamefont
  {Carballeira}}, \bibinfo {author} {\bibfnamefont {J.}~\bibnamefont
  {Mosqueira}}, \bibinfo {author} {\bibfnamefont {A.}~\bibnamefont
  {Revcolevschi}}, \ and\ \bibinfo {author} {\bibfnamefont {F.}~\bibnamefont
  {Vidal}},\ }\bibfield  {title} {\emph {\bibinfo {title} {{First observation
  for a cuprate superconductor of fluctuation-induced diamagnetism well inside
  the finite-magnetic-field regime}}},\ }\href {\doibase
  10.1103/PhysRevLett.84.3157} {\bibfield  {journal} {\bibinfo  {journal}
  {Physical Review Letters}\ }\textbf {\bibinfo {volume} {84}},\ \bibinfo
  {pages} {3157} (\bibinfo {year} {2000})}\BibitemShut {NoStop}%
\bibitem [{\citenamefont {Pope}\ \emph {et~al.}(2001)\citenamefont {Pope},
  \citenamefont {{Littleton IV}},\ and\ \citenamefont {Tritt}}]{Pope2001a}%
  \BibitemOpen
  \bibfield  {author} {\bibinfo {author} {\bibfnamefont {A.~L.}\ \bibnamefont
  {Pope}}, \bibinfo {author} {\bibfnamefont {R.~T.}\ \bibnamefont {{Littleton
  IV}}}, \ and\ \bibinfo {author} {\bibfnamefont {T.~M.}\ \bibnamefont
  {Tritt}},\ }\bibfield  {title} {\emph {\bibinfo {title} {{Apparatus for the
  rapid measurement of electrical transport properties for both "needle-like"
  and bulk materials}}},\ }\href {\doibase 10.1063/1.1380390} {\bibfield
  {journal} {\bibinfo  {journal} {Review of Scientific Instruments}\ }\textbf
  {\bibinfo {volume} {72}},\ \bibinfo {pages} {3129} (\bibinfo {year}
  {2001})}\BibitemShut {NoStop}%
\bibitem [{\citenamefont {{Quantum Design}}(1999)}]{QuantumDesign1999}%
  \BibitemOpen
  \bibfield  {author} {\bibinfo {author} {\bibnamefont {{Quantum Design}}},\
  }\bibfield  {title} {\emph {\bibinfo {title} {{Physical Property Measurement
  System: Resistivity Option User ' s Manual}}},\ }\href@noop {} {\bibfield
  {journal} {\bibinfo  {journal} {PPMS Manual}\ } (\bibinfo {year}
  {1999})}\BibitemShut {NoStop}%
\end{thebibliography}

%

\end{document}